\documentclass[sigconf, nonacm]{acmart}

\newcommand\vldbdoi{10.14778/3797919.3797942}
\newcommand\vldbpages{1386 - 1399}
\newcommand\vldbvolume{19}
\newcommand\vldbissue{6}
\newcommand\vldbyear{2026}
\newcommand\vldbauthors{\authors}
\newcommand\vldbtitle{\shorttitle} 
\newcommand\vldbavailabilityurl{https://github.com/ruc-datalab/TACO-Benchmark}
\newcommand\vldbpagestyle{empty}

\makeatletter
\def\@ACM@checkaffil{
    \if@ACM@instpresent\else
    \ClassWarningNoLine{\@classname}{No institution present for an affiliation}%
    \fi
    \if@ACM@citypresent\else
    \ClassWarningNoLine{\@classname}{No city present for an affiliation}%
    \fi
    \if@ACM@countrypresent\else
        \ClassWarningNoLine{\@classname}{No country present for an affiliation}%
    \fi
}

\makeatother

\renewcommand{\abstractname}{ABSTRACT}
\renewcommand{\refname}{REFERENCES}

\usepackage[a-2b]{pdfx}

\usepackage{color}
\usepackage{xcolor}
\usepackage{longtable}
\usepackage{multirow}
\usepackage{xspace}
\usepackage{booktabs}
\usepackage{graphicx}
\usepackage{amsfonts}
\usepackage{algorithm}
\usepackage{algorithmicx}
\usepackage{algpseudocode}
\usepackage{algcompatible}
\usepackage{amsmath}
\usepackage{enumitem}
\usepackage{booktabs}
\usepackage{caption}
\usepackage{pifont}
\usepackage{listings}
\usepackage{makecell}
\usepackage[normalem]{ulem}
\usepackage{float}

\usepackage{fancyhdr}

\usepackage{tcolorbox}
\tcbuselibrary{breakable}
\definecolor{dc}{RGB}{241, 158, 194}
\AtBeginDocument{%
  }

\newcommand{\blue}[1]{\textcolor{black}{#1}}

\newcommand{\sys}{\textsc{TACO}\xspace}
\newcommand{\model}{\textsc{TACO-SQL}\xspace}

\newcommand{\term}[1]{{\tt #1}}

\newcommand{\nlsql}{Text-to-SQL\xspace}

\newcommand{\sstab}{\rule{0pt}{8pt}\\[-2.2ex]}

\newcommand{\stitle}[1]{\sstab\noindent{\bf #1}}
\newcommand{\etitle}[1]{\vspace{1mm}\noindent{\underline{\em #1}}}

\begin{document}

\title{\sys: A Benchmark for Open-Domain Text-to-SQL with Ambiguous and Cross-Database Queries}


%
\author{Chao Deng}
\affiliation{
  \institution{Renmin University, China}
}
\email{dengc@ruc.edu.cn}

\author{Ju Fan}
\authornote{Ju Fan and Xiaofeng Jia are the corresponding authors.}
\affiliation{%
  \institution{Renmin University, China}
}
\email{fanj@ruc.edu.cn}

\author{Yuyu Luo}
\affiliation{%
  \institution{HKUST (GZ) / HKUST}
}
\email{yuyuluo@hkust-gz.edu.cn}

\author{Qinliang Xue}
\affiliation{%
  \institution{Renmin University, China}
}
\email{xueql@ruc.edu.cn}

\author{Meihao Fan}
\affiliation{%
  \institution{Renmin University, China}
}
\email{fmh1art@ruc.edu.cn}

\author{Yuxin Zhang}
\affiliation{%
  \institution{Renmin University, China}
}
\email{yuxin.zhang@ruc.edu.cn}

\author{Min Zhang}
\affiliation{%
  \institution{Renmin University, China}
}
\affiliation{%
  \institution{Beijing Big Data Centre}
}
\email{zmbbdc@gmail.com}

\author{Xiaofeng Jia}
\authornotemark[1]
\affiliation{%
  \institution{Beijing Big Data Centre}
}
\email{jiaxiaofeng119@gmail.com}

\author{Jing Zhang}
\affiliation{%
  \institution{Renmin University, China}
}
\email{zhang-jing@ruc.edu.cn}

\author{Xiaoyong Du}
\affiliation{%
  \institution{Renmin University, China}
}
\email{duyong@ruc.edu.cn}

\begin{abstract}
\label{sec:abstract}

\nlsql aims to translate natural language questions into executable SQL queries over structured databases. Existing benchmarks mainly focus on closed-domain settings with predefined database schemas and well-specified questions, but they fall short in addressing the challenges of open-domain scenarios, such as ambiguous questions, unspecified databases, and cross-database querying.
To bridge this gap, we introduce \sys, a benchmark for open-domain \underline{T}ext-to-SQL with \underline{A}mbiguous and \underline{C}r\underline{O}ss-database queries. \sys consists of 1,500 real-world \nlsql examples based on a smart city data service and 13,000 high-quality synthetic examples generated based on large-scale open data portals, covering diverse domains such as transportation, healthcare, and finance. To construct the synthetic examples, we develop an effective data synthesis pipeline that preserves the complexity of real-world queries.
To demonstrate the utility of \sys, we introduce a baseline \sys-SQL composed of question rewriting, table linking, and query planning, to illustrate the challenges posed by TACO and to better understand the limitations of existing \nlsql approaches. Extensive experiments on \sys using a variety of recent \nlsql approaches show that, while \sys-SQL achieves the best results, a significant gap still remains between the existing approaches and human-written SQL. These findings highlight the difficulty of open-domain \nlsql and position \sys as a valuable benchmark to drive future research.

\end{abstract}

\maketitle
\thispagestyle{empty}
\pagestyle{\vldbpagestyle}
\begingroup\small\noindent\raggedright\textbf{PVLDB Reference Format:}\\
\vldbauthors. 
\vldbtitle. PVLDB, \vldbvolume(\vldbissue): \vldbpages, \vldbyear.\\
\href{https://doi.org/\vldbdoi}{doi:\vldbdoi}
\endgroup
\begingroup
\renewcommand\thefootnote{}\footnote{\noindent
This work is licensed under the Creative Commons BY-NC-ND 4.0 International License. Visit \url{https://creativecommons.org/licenses/by-nc-nd/4.0/} to view a copy of this license. For any use beyond those covered by this license, obtain permission by emailing \href{mailto:info@vldb.org}{info@vldb.org}. Copyright is held by the owner/author(s). Publication rights licensed to the VLDB Endowment. \\
\raggedright Proceedings of the VLDB Endowment, Vol. \vldbvolume, No. \vldbissue\ %
ISSN 2150-8097. \\
\href{https://doi.org/\vldbdoi}{doi:\vldbdoi} \\
}\addtocounter{footnote}{-1}\endgroup

\ifdefempty{\vldbavailabilityurl}{}{
\vspace{.3cm}
\begingroup\small\noindent\raggedright\textbf{PVLDB Artifact Availability:}\\
The source code, data, and/or other artifacts have been made available at \url{\vldbavailabilityurl}.
\endgroup
}

\section{INTRODUCTION}
\label{sec:Introduction}

\nlsql, which translates natural language (NL) questions into SQL queries, provides a user-friendly interface for non-technical users to access structured data, supporting applications in business intelligence and data-driven decision-making~\cite{10.14778/3401960.3401970,qin2022surveytexttosqlparsingconcepts,yu2019spiderlargescalehumanlabeleddataset,deng2022recentadvancestexttosqlsurvey,DBLP:journals/corr/abs-2510-23587,zhu2025elliesql}. 
With the rapid advancement of large language models (LLMs), many LLM-powered approaches have emerged (see the recent survey~\cite{DBLP:journals/tkde/LiuSLMJZFLTL25}), achieving strong performance on standard \nlsql benchmarks~\cite{10.14778/3401960.3401970, qin2022surveytexttosqlparsingconcepts, liu2025surveynl2sqllargelanguage,liu2025nl2sqlbugsbenchmarkdetectingsemantic}.

\begin{figure*}[t]
	\centering    \includegraphics[width=1\textwidth]{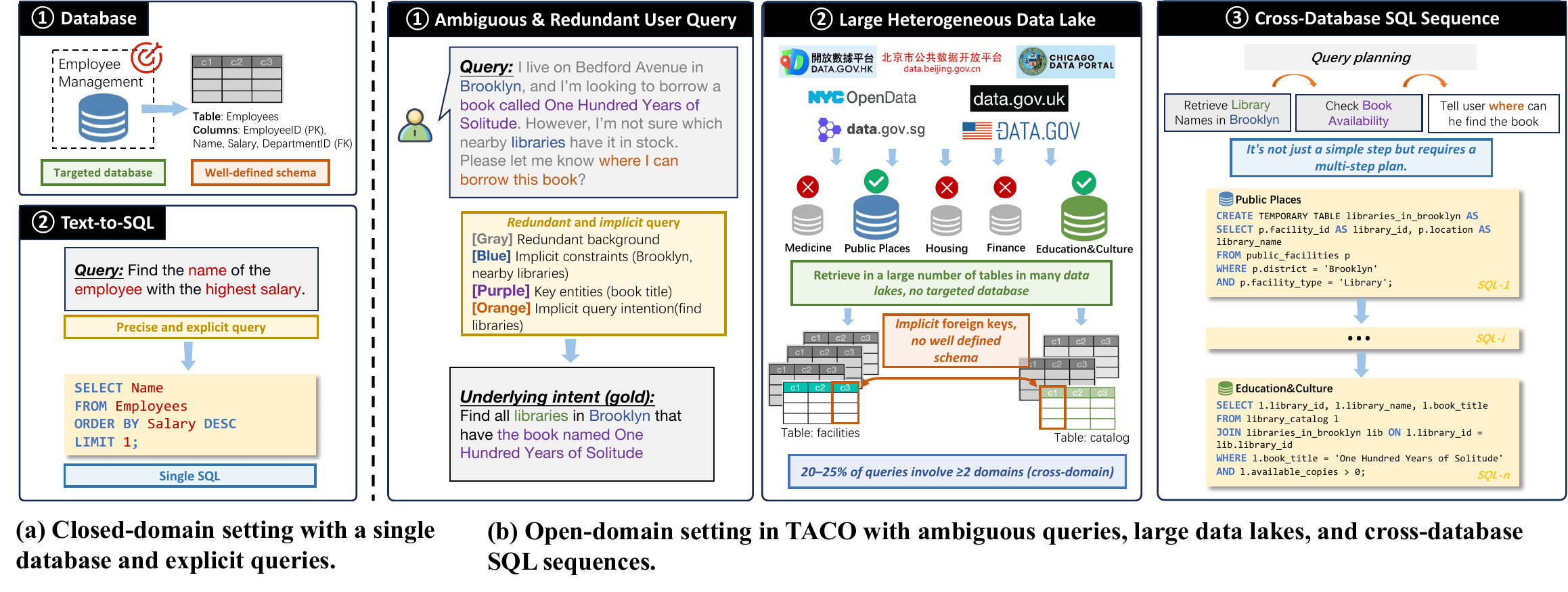}
    \caption{\blue{Comparison of closed-domain and open-domain \nlsql scenarios.}}
	\label{fig:open-domain-nlsql-scenarios}
    \vspace{-1em}
\end{figure*}

\stitle{Traditional Focus: Closed-Domain Text-to-SQL.} 
Most prior work focuses on the \emph{closed-domain} setting, as shown in Figure~\ref{fig:open-domain-nlsql-scenarios}(a), where the target database is predefined and the schema is fully observable. Users pose relatively precise questions, and the task reduces to translating an NL question into a single SQL query over a known schema.
For example, the question ``\emph{Find the name of the employee with the highest salary}'' can be translated into an SQL query given the schema of an \texttt{Employee} \texttt{Management} database.

\stitle{New Challenges: Open-Domain \nlsql.} 
In many real deployments, data is organized as large \emph{data lakes}, such as smart-city data services~\cite{camero2019smartcity} and open government data portals~\cite{beijingdata, usdata}, where information spans numerous heterogeneous databases. 
In these settings, the closed-domain assumptions of \nlsql often do not hold~\cite{Li_2024}, and three challenges arise:

  
\textbf{(1) Ambiguous NL questions.}
  Users often have limited knowledge of the underlying data and issue vague or redundant questions with unclear intent or incomplete constraints~\cite{hazoom2021texttosqlwildnaturallyoccurringdataset, chen2025beaverenterprisebenchmarktexttosql, saparina2024ambrosiabenchmarkparsingambiguous, wang-etal-2023-know, bhaskar2023benchmarkingimprovingtexttosqlgeneration}.
  
   \textbf{(2) Unspecified target databases.}
  NL Questions rarely specify the relevant databases or tables; the system must retrieve candidate tables from a large, heterogeneous data lake~\cite{zhang2024murremultihoptableretrieval,chen-etal-2024-table, 10.1145/3709727}.
  
  \textbf{(3) Cross-database querying.}
  Answering a single NL question may require combining data from multiple databases with weak or implicit relationships, requiring multi-step query planning and result integration~\cite{guo2019complextexttosqlcrossdomaindatabase,yu2019sparccrossdomainsemanticparsing}.

\begin{example}
	Figure~\ref{fig:open-domain-nlsql-scenarios}(b) illustrates an open-domain scenario over an open data portal such as \texttt{NYC} \texttt{Open} \texttt{Data}\footnote{https://opendata.cityofnewyork.us/}. 
	A user may ask: \emph{``Could you tell me which libraries around Brooklyn have a copy of \textit{One Hundred Years of Solitude}?''} 
	The question is informally phrased and omits schema details, and relevant information may reside in separate databases(e.g., \texttt{Public} \texttt{Places} and \texttt{Education} \texttt{\&} \texttt{Culture}). Thus, answering the question requires a sequence of SQL queries whose intermediate results must be joined across databases. 
	This combination of ambiguity, database discovery, and cross-database reasoning extends beyond traditional \nlsql benchmarks.
\end{example}

\stitle{Limitations of Existing Benchmarks.}
Several benchmarks have advanced \nlsql research, including WikiSQL~\cite{zhong2017seq2sqlgeneratingstructuredqueries}, Spider~\cite{yu2019spiderlargescalehumanlabeleddataset}, BIRD~\cite{li2024can}, Spider 2.0~\cite{lei2024spider20evaluatinglanguage}, and NL2SQL-BUGs~\cite{liu2025nl2sqlbugsbenchmarkdetectingsemantic}. 
These benchmarks typically assume a single target database (or a small set of databases) with fully observable schemas and relatively concise NL questions, and thus primaryly evaluate closed-domain settings. 
Recent diagnostic datasets (e.g., SNAILS~\cite{10.1145/3709727}, Ambrosia~\cite{saparina2024ambrosiabenchmarkparsingambiguous}) consider schema naming and ambiguity, but still operate under the same closed-domain assumption.
Closer to our setting, union-style and interaction benchmarks such as Bird-Union/Spider-Union~\cite{vaidya2025tailorsql}, BIRD-Interact~\cite{huo2025birdinteract}, and SWE-SQL (Bird-Critic)~\cite{li2025swesql} address challenges of schema scale, multi-turn clarification, or SQL debugging. However, they generally rely on a unified schema or reuse existing benchmarks, and do not model data-lake table retrieval and cross-database execution over heterogeneous datasets.
Therefore, a benchmark that jointly evaluates ambiguous questions, large-scale table retrieval, and cross-database reasoning remains missing.

\stitle{Our Open-Domain Text-to-SQL Benchmark: \sys.}
To address this gap, we introduce \sys, a benchmark for open-domain \underline{T}ext-to-SQL with \underline{A}mbiguous and \underline{C}r\underline{O}ss-database queries.
\sys evaluates three capabilities of \nlsql approaches: 
(i) resolving ambiguity and redundancy in user NL questions, 
(ii) retrieving relevant tables from large heterogeneous data lakes, and 
(iii) planning and generating SQL pipelines that may span multiple databases.
\sys consists of two complementary datasets, each consisting of NL questions paired with gold SQL (or short SQL sequences):

{\textbf{Type-1 (\sys-SmartCity)}: Real Examples from Smart City Services.}
We collect 1,500 NL questions from a Beijing smart city data service, where data from 31 government departments are integrated into a shared data lake.
Domain experts decompose each question, annotate executable SQLs, and validate results via cross-execution checking.
The resulting benchmark subset covers 117 tables across 31 databases, with an average of 1.19 databases per query, and includes many cases with vague intent, implicit constraints, and cross-database reasoning.

\emph{\textbf{Type-2 (\sys-OpenData with \sys-Beijing and \sys-US)}: Synthetic Examples from Open Data Portals.}
Real-world data is limited in scale and constrained by privacy. 
To enable large-scale, fully public evaluation, we develop a data synthesis pipeline that preserves structural patterns of real SQLs, populates content using public schema, and generates NL questions via Chain-of-Thought prompting. 
Applying this pipeline to the Beijing Municipal Open Data Platform~\cite{beijingdata} and the U.S. Government’s Open Data Portal~\cite{usdata} produces 13{,}000 bilingual examples across 52 databases and 13{,}004 tables, forming the \sys-Beijing and \sys-US datasets.

\begin{table*}[t!]
  \centering
  \vspace{-1em} 
  \caption{\bf{Comparison of \sys with representative \nlsql benchmarks, including recent ambiguity-, interaction-, and union-style benchmarks.}}
  \vspace{-2mm}
  \label{tab:nl2sql_benchmarks}
  \begin{tabular}{|cc||c|c|c|c||c|c|c|c|}
  \hline
  \multicolumn{2}{|c||}{Datasets} 
    & {\#-Examples} 
    & {\#-DBs} 
    & {\#-Tables} 
    & {\#-Tables/DB} 
    & Redundancy 
    & Table linking 
    & Cross Table 
    & Cross DB \\ \hline\hline

  \multicolumn{2}{|c||}{WikiSQL~\cite{zhong2017seq2sqlgeneratingstructuredqueries}} 
    & 80{,}654 & 26{,}521 & 26{,}521 & 1.0 
    & $\times$ & $\times$ & $\times$ & $\times$ \\ \hline

  \multicolumn{2}{|c||}{Spider1.0~\cite{yu2019spiderlargescalehumanlabeleddataset}} 
    & 10{,}181 & 200 & 1{,}020 & 5.1 
    & $\times$ & $\times$ & \checkmark & $\times$ \\ \hline

  \multicolumn{2}{|c||}{KaggleDBQA~\cite{lee2021kaggledbqarealisticevaluationtexttosql}} 
    & 272 & 8 & 18 & 2.25 
    & $\times$ & $\times$ & \checkmark & $\times$ \\ \hline

  \multicolumn{2}{|c||}{BIRD~\cite{li2024can}} 
    & 12{,}751 & 95 & 694 & 7.3 
    & $\times$ & $\times$ & \checkmark & $\times$ \\ \hline

  \multicolumn{2}{|c||}{Spider2.0-snow~\cite{lei2024spider20evaluatinglanguage}} 
    & 547 & 152 & 8{,}000 & 52.63 
    & $\times$ & $\times$ & \checkmark & $\times$ \\ \hline

\multicolumn{2}{|c||}{\blue{Bird-Critic~\cite{li2025swesql}}} 
    & \blue{--} & \blue{95} & \blue{694} & \blue{7.3} 
    & \blue{\checkmark} & \blue{\checkmark} & \blue{\checkmark} & \blue{$\times$} \\ \hline

\multicolumn{2}{|c||}{\blue{BIRD-Interact~\cite{huo2025birdinteract}}} 
    & \blue{6{,}500} & \blue{95} & \blue{694} & \blue{7.3} 
    & \blue{\checkmark} & \blue{\checkmark} & \blue{\checkmark} & \blue{$\times$} \\ \hline

\multicolumn{2}{|c||}{\blue{SpiderUnion~\cite{vaidya2025tailorsql}}} 
    & \blue{13{,}000} & \blue{600} & \blue{5{,}000} & \blue{8.3} 
    & \blue{$\times$} & \blue{$\times$} & \blue{\checkmark} & \blue{$\times$} \\ \hline

\multicolumn{2}{|c||}{\blue{BirdUnion~\cite{vaidya2025tailorsql}}} 
    & \blue{22{,}000} & \blue{200} & \blue{3{,}000} & \blue{15.0} 
    & \blue{$\times$} & \blue{\checkmark} & \blue{\checkmark} & \blue{$\times$} \\ \hline

\multicolumn{2}{|c||}{\blue{Abacus-SQL~\cite{xu2025abacussql}}} 
    & \blue{--} & \blue{--} & \blue{--} & \blue{--} 
    & \blue{$\times$} & \blue{--} & \blue{\checkmark} & \blue{\checkmark} \\ \hline\hline

  \multicolumn{1}{|c|}{\multirow{3}{*}{TACO}} 
    & TACO-SmartCity 
    & 1{,}500 & 31 & 113 & 3.77 
    & \checkmark & \checkmark & \checkmark & \checkmark \\ \cline{2-10} 
  \multicolumn{1}{|c|}{} 
    & TACO-Beijing 
    & 7{,}000 & 28 & 3{,}010 & 107.50 
    & \checkmark & \checkmark & \checkmark & \checkmark \\ \cline{2-10} 
  \multicolumn{1}{|c|}{} 
    & TACO-US 
    & 6{,}000 & 24 & 9{,}994 & 416.42 
    & \checkmark & \checkmark & \checkmark & \checkmark \\ \hline
  \end{tabular}
\end{table*}

\begin{figure}[t!]
	\centering
	\includegraphics[width=\columnwidth]{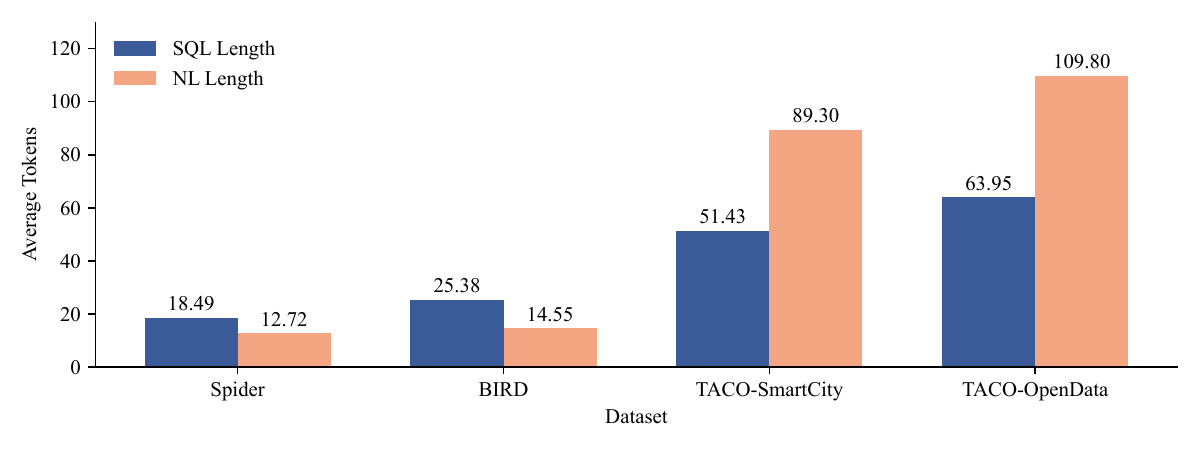}
	\caption{Average tokens in SQL and NL across benchmarks.}
	\label{fig:average_token_length}
    \vspace{-1em}
\end{figure}

The real and synthetic datasets form a unified testbed for open-domain \nlsql. 
As shown in Table~\ref{tab:nl2sql_benchmarks}, \sys matches or exceeds classic benchmarks in the number of databases and tables, and explicitly targets large, heterogeneous data lakes. 
Furthermore, compared with Spider and BIRD, \sys contains substantially longer NL questions and SQL queries (Figure~\ref{fig:average_token_length}), indicating higher levels of ambiguity, redundancy, and structural complexity.

\stitle{Remarks.}
\sys adopts a \emph{pure SQL–based execution pipeline}, ensuring that the benchmark evaluates Text-to-SQL reasoning rather than external scripting. Specifically, each open-domain NL question is mapped to a \emph{logical SQL pipeline} composed of multiple SQL statements, where each statement is executed on its corresponding database. Intermediate results are combined \emph{within the database engine} using standard SQL constructs (e.g., \texttt{JOIN}, \texttt{UNION}, nested subqueries), without relying on Python or shell-level orchestration. 

\stitle{Our Open-Domain Text-to-SQL Baseline: \sys-SQL.}
In order to demonstrate the utility of \sys and analyze existing \nlsql approaches, we introduce \sys-SQL, a modular LLM-based framework for open-domain \nlsql.
It consists of three components:
(1) \emph{Question Rewriting}, which reformulates ambiguous NL questions to clarify user intent; 
(2) \emph{Table Linking}, which retrieves relevant tables from large heterogeneous data lakes; and 
(3) \emph{Query Planning and Generation}, which decomposes complex queries into multi-step plans and produces executable SQL pipelines.
Note that the primary contribution of this work is the \sys benchmark. \sys-SQL only serves as a representative baseline that instantiates these components and reveals where the existing models succeed or fail under open-domain conditions.

We conduct an extensive empirical evaluation on \sys, comparing a diverse set of \nlsql approaches, including base LLMs (e.g., GPT series, Llama 3, DeepSeek), LLM-based systems (e.g., DIN-SQL~\cite{pourreza2023dinsqldecomposedincontextlearning}, MAC-SQL~\cite{wang2024macsqlmultiagentcollaborativeframework}), supervised fine-tuned models (e.g., CodeS~\cite{10.1145/3654930}, Qwen2.5-Coder), and hybrid strategies (e.g., CHESS~\cite{talaei2024chesscontextualharnessingefficient}, ZeroNL2SQL~\cite{10.14778/3681954.3681960}). 
The results show that \sys-SQL consistently improves over the LLM-based baselines, yet a substantial gap remains relative to human experts, indicating the difficulty of open-domain \nlsql.
%

\stitle{Contributions.}
Our main contributions are as follows:

(1) We formalize the open-domain \nlsql problem (Section~\ref{sec:pre}) and introduce \sys, a benchmark designed to evaluate ambiguity resolution, data-lake table retrieval, and cross-database SQL reasoning (Section~\ref{sec:dataset-construction}). 

(2) We propose \sys-SQL, a LLM-based framework that integrates question rewriting, table linking, and query planning, serving as a practical baseline for open-domain \nlsql (Section~\ref{sec:Framework}).

(3) We conduct experiments on \sys using a diverse set of \nlsql methods (Section~\ref{sec:Experiments}).
The results show a substantial performance gap between the approaches and gold SQL, indicating that \sys poses a challenging benchmark for advancing open-domain \nlsql research.
%

\section{PRELIMINARIES}
\label{sec:pre}
This section first formalizes the open-domain \nlsql problem and its benchmark, and then reviews related work.

\subsection{Problem Formulation}
\label{subsec:problem formulation}

\stitle{Open-Domain \nlsql.}
The open-domain \nlsql task aims to map a natural language (NL) question \(Q\) to one or more SQL queries \(S=\{S_1,\ldots,S_k\}\), defined as
\( S = g(Q, \mathcal{D} \mid \Phi) \),
where \(g\) denotes a model with parameters \(\Phi\), and
\(\mathcal{D}=\{DB_1,\ldots,DB_n\}\) is a collection of heterogeneous databases.
Each database \(DB_i\) contains tables \(T\) and columns \(C\), i.e.,  
\( DB_i=\langle T, C\rangle \).
The generated SQL queries may form an ordered sequence: each \(S_i\) specifies a target database and may consume intermediate results produced by earlier steps, yielding a logical execution pipeline whose final result is the query answer.

Open-domain queries typically exhibit:  
(1) ambiguous or redundant phrasing;  
(2) missing or implicit database/table references; and  
(3) information distributed across multiple heterogeneous databases.  
Figure~\ref{fig:open-domain-nlsql-scenarios}(b) illustrates such a setting over open-data portals.

\stitle{Types of Supported Queries: }
In \sys, queries are categorized by the origin of the tables they involve. The current version of \sys supports three types of \nlsql queries:
\textbf{(1)} single-table selection and \textbf{(2)} multi-table join queries within a {single database}, and \textbf{(3)} {cross-database} queries involving tables from multiple databases.
For cross-database queries, intermediate results are combined \emph{purely through SQL} (e.g., \texttt{JOIN}, \texttt{UNION}, nested subqueries) within the database engine, without Python or shell-level scripting.


\stitle{Open-Domain \nlsql Benchmark.}
We formalize an open-domain benchmark as \((\mathcal{D},\{Q,S\})\), where \(\mathcal{D}\) is a collection of heterogeneous databases and each example consists of an NL question and its corresponding SQL queries.
Note that \sys focuses on single-turn \nlsql: each example contains one NL question and one executable SQL query (or SQL sequence) that runs directly on~\(\mathcal{D}\).
The benchmark does not include multi-turn dialogue, agentic data-analysis pipelines (e.g., Python/UDF workflows), or non-SQL tasks such as data cleaning or schema repair.
Although some queries may require schema- or value-level background knowledge, all answers are obtained strictly through SQL execution over the underlying databases.

\subsection{Related Work}
\label{subsec:related_work}

\stitle{\nlsql.}
LLM-based approaches have substantially advanced Text-to-SQL through in-context learning, instruction tuning, and code generation~\cite{Li_2024,DBLP:journals/pvldb/LuoLFCT25,DBLP:journals/corr/abs-2505-07437,wu2026autowebworldsynthesizinginfiniteverifiable}. For example, fine-tuned models like CodeS~\cite{10.1145/3654930} directly enhance generation capabilities. More advanced frameworks adopt agentic strategies~\cite{ruan2026aorchestraautomatingsubagentcreation,DBLP:conf/iclr/ZhangXYTCCZCHWZ25}. For example, Alpha-SQL~\cite{li2025alphasqlzeroshottexttosqlusing} introduces a test-time scaling method~\cite{DBLP:conf/emnlp/XiangZYLTTRTHWL25} that leverages Monte Carlo Tree Search (MCTS) to iteratively explore the SQL generation space. CHASE-SQL~\cite{pourreza2024chasesqlmultipathreasoningpreference} employs multi-agent collaboration. Hybrid systems combine LLM reasoning with specialized modules; examples include CHESS~\cite{talaei2024chesscontextualharnessingefficient} and ZeroNL2SQL~\cite{10.14778/3681954.3681960}, which integrate schema linking and self-refinement~\cite{zhang2025rewardsqlboostingtexttosqlstepwise, li2025alphasqlzeroshottexttosqlusing}. Furthermore, system-level pipelines for open-domain Text-to-SQL incorporate retrieval, rewriting, and debugging~\cite{xu2025abacussql}. A notable example is DeepEye-SQL~\cite{DBLP:journals/corr/abs-2510-17586}, which formulates the task as a software engineering process with N-version generation and multi-step verification. Despite their strong performance in closed-domain settings, these methods do not explicitly address ambiguity, unspecified schemas, and cross-database planning, thus motivating the need for an open-domain benchmark~\cite{ma2024plugandplaynaturallanguagerewriter}.

\stitle{Benchmarks for \nlsql.}
Early benchmarks such as \emph{WikiSQL}~\cite{zhong2017seq2sqlgeneratingstructuredqueries} evaluate simple single-table queries.  
\emph{Spider}~\cite{yu2019spiderlargescalehumanlabeleddataset} and \emph{BIRD} introduce complex multi-table queries across diverse databases, and \emph{Spider2.0}~\cite{lei2024spider20evaluatinglanguage} further extends coverage to enterprise-level SQL patterns.
Diagnostic benchmarks like \emph{Dr.Spider}~\cite{chang2023drspiderdiagnosticevaluationbenchmark} and \emph{NL2SQL-BUGs}~\cite{liu2025nl2sqlbugsbenchmarkdetectingsemantic} evaluate dataset quality and semantic robustness.
Other benchmarks target specific subproblems:
SNAILS~\cite{10.1145/3709727} focuses on schema linking,  
Ambrosia~\cite{saparina2024ambrosiabenchmarkparsingambiguous} and Wang et al.~\cite{wang-etal-2023-know} study ambiguity and unanswerable questions, and 
LogicalBeam~\cite{bhaskar2023benchmarkingimprovingtexttosqlgeneration} explores program search but does not consider open-domain table retrieval.

Closer to our setting, \emph{TailorSQL}~\cite{vaidya2025tailorsql} introduces \emph{Spider-Union} and \emph{Bird-Union}, which merge schemas from Spider/BIRD into large unified databases to evaluate join-path prediction and table selection.  
However, each query still operate within a single logical database, and SQL queries are derived from existing benchmarks rather than real user inputs.
\emph{BIRD-INTERACT}~\cite{huo2025birdinteract} evaluates \nlsql in multi-turn conversational and agentic settings by augmenting BIRD with hierarchical knowledge bases and user simulators.  
\emph{SWE-SQL} (also known as \emph{BIRD-CRITIC})~\cite{li2025swesql} provides a diagnostic benchmark for debugging user-written SQL rather than mapping NL to SQL.

While these benchmarks expand coverage of robustness, ambiguity, interaction, and workload variation, they do not model the open-domain setting (e.g., municipal or open-data platforms), where NL questions may be ambiguous, the target database is unspecified, and relevant information can span multiple heterogeneous databases.
\sys addresses this setting by grounding the benchmark in real user queries and by explicitly evaluating data-lake table retrieval and cross-database SQL reasoning.
%

\section{THE \sys BENCHMARK}

\label{sec:dataset-construction}

To advance research on open-domain \nlsql, we introduce \sys, a benchmark for \underline{T}ext-to-SQL with \underline{A}mbiguous and \underline{C}r\underline{O}ss-database queries.
\sys consists of two complementary parts:
(1) \sys-SmartCity, comprising 1{,}500 real-world NL--SQL examples from a municipal data service; and  
(2) \sys-OpenData (Beijing/US), containing 13{,}000 synthetic examples constructed from large-scale open-data portals.
Each example pairs an NL question with an executable SQL query (or an SQL sequence) that can be directly executed over the underlying heterogeneous databases. 
We describe the curation of real examples in Section~\ref{subsec:real-examples}, the synthesis pipeline in Section~\ref{subsec:synthetic-examples}, and benchmark statistics in Section~\ref{subsec:benchmark-statistics}.

\subsection{Curating Real \nlsql Examples}
\label{subsec:real-examples}

To support evaluation on real-world workloads in open-domain settings, we curate real NL-SQL pairs from a Smart City Data Service.
The environment integrates databases from multiple departments, yet users typically pose questions without specifying the target database or table, leading to ambiguity, implicit schema reference, and cross-database reasoning requirements.
Note that all \sys-SmartCity examples are fully anonymized, with no personal identifiers retained; only the NL-SQL pairs are used for benchmarking.

\begin{figure}
    \centering    \includegraphics[width=0.5\textwidth]{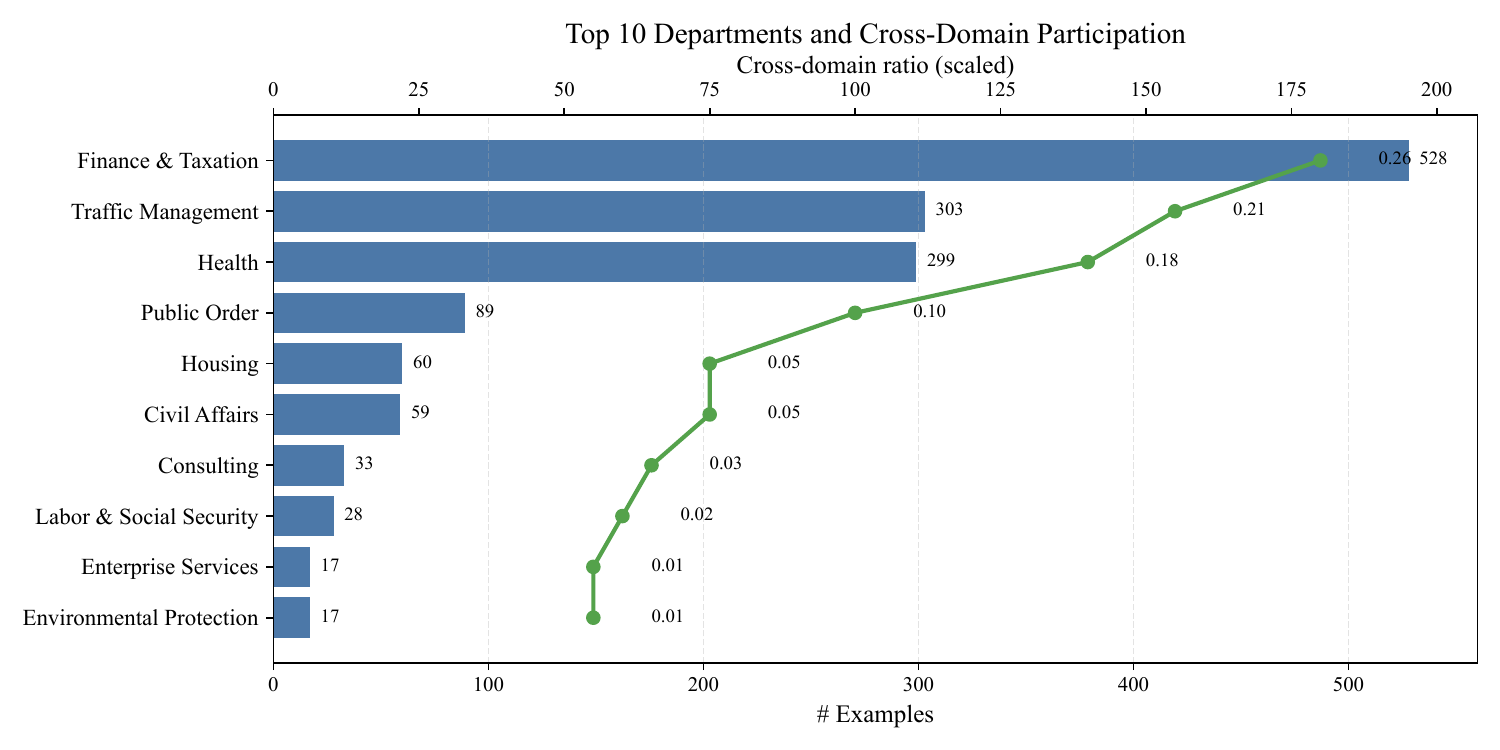}
    \vspace{-1em}
    \caption{Question distribution by department in the Smart City Data Service dataset (top-10 departments). Bars show the number of queries involving each department, and the line plot shows the proportion of cross-database queries (i.e., queries involving this department and at least one additional department/database).
}
    \label{fig:10departments}
\end{figure}

\begin{figure*}
    \centering
    \includegraphics[width=0.88\textwidth]{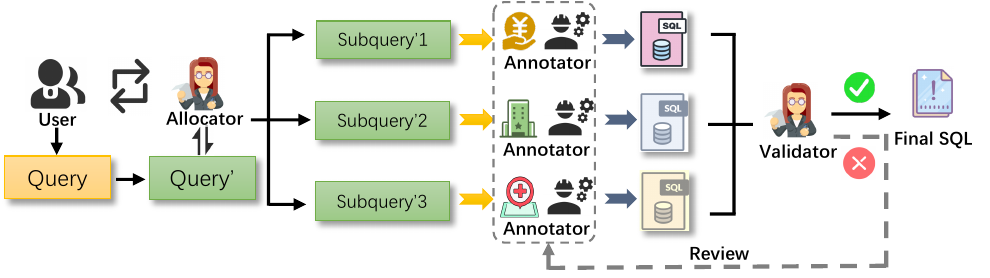}
    \caption{Curation workflow for real \nlsql examples.
    }
    
    \label{fig:dataset1}
\end{figure*}

The Smart City Data Service in Beijing integrates data from 31 government departments. User questions, collected from online platforms, hotlines, and offline service centers, often mention domain entities (e.g., district names) but rarely specify the target database, making ambiguity and schema discovery central challenges. Figure~\ref{fig:10departments} shows the distribution of queries across departments.


\begin{example}[\blue{Real-World Data: Ambiguous, Cross-Department Query}]
\label{ex:smartcity-ambiguous}
\blue{A user reports an auto repair shop in an industrial zone that is ``causing pollution and appears to operate without proper licenses'', and requests to \emph{``check environmental monitoring data for businesses in this area to verify compliance with emission standards''}. The question references environmental monitoring and business registration but does not specify the relevant databases, tables, or precise locations. 
The ground-truth SQL is:}
\begin{verbatim}
SELECT monitortime, so2, no3
FROM environmental_bureau.enterprise_emission
WHERE corporationcode IN (
  SELECT ent_name
  FROM market_supervision.business_registration_info
  WHERE business_address LIKE '%Industrial Zone%'
);
\end{verbatim}
\end{example}

\stitle{Curation Workflow.}
Our curation pipeline (Figure~\ref{fig:dataset1}) follows a multi-stage, expert-driven process.

\etitle{(1) Query Decomposition.}
A centralized allocator reviews each user question and manually decomposes complex queries into simpler subqueries while preserving the original semantics.

\etitle{(2) Expert Annotation.}
Each subquery is assigned to domain experts who author the corresponding executable SQL.
No LLMs or automatic rewriting components are involved; all SQL annotations in \sys-SmartCity are written entirely by human experts.

\etitle{(3) Validation and Integration.}
A separate validation team executes each SQL query on the databases and verifies the correctness.
For multi-step queries, intermediate results are combined to produce the final answer.
All result integration is performed using \emph{SQL-only} operations (e.g., \texttt{JOIN}, \texttt{UNION}, and nested subqueries) within the database engine, without Python or shell-level orchestration.
Each NL-SQL pair is accepted only if:  
(i) the SQL executes successfully and returns a correct result verified by experts; and 
(ii) the allocator confirms that the combined result answers the original user question.  
%
From this multi-stage pipeline, we select 1{,}500 high-quality examples whose SQL queries correctly answer the user NL questions, forming the real-world dataset of \sys.

\stitle{Remarks.}
Although all SmartCity queries are anonymized, access follows standard practices for restricted datasets. Specifically, qualified researchers may either \emph{apply for controlled access} via a data-use agreement or \emph{submit their models to us for evaluation}. Submitted models are executed in an isolated environment, and only aggregated metrics are returned. 

\subsection{Synthesizing \nlsql Examples}
\label{subsec:synthetic-examples}

\begin{figure*}
    \centering    \includegraphics[width=1\textwidth]{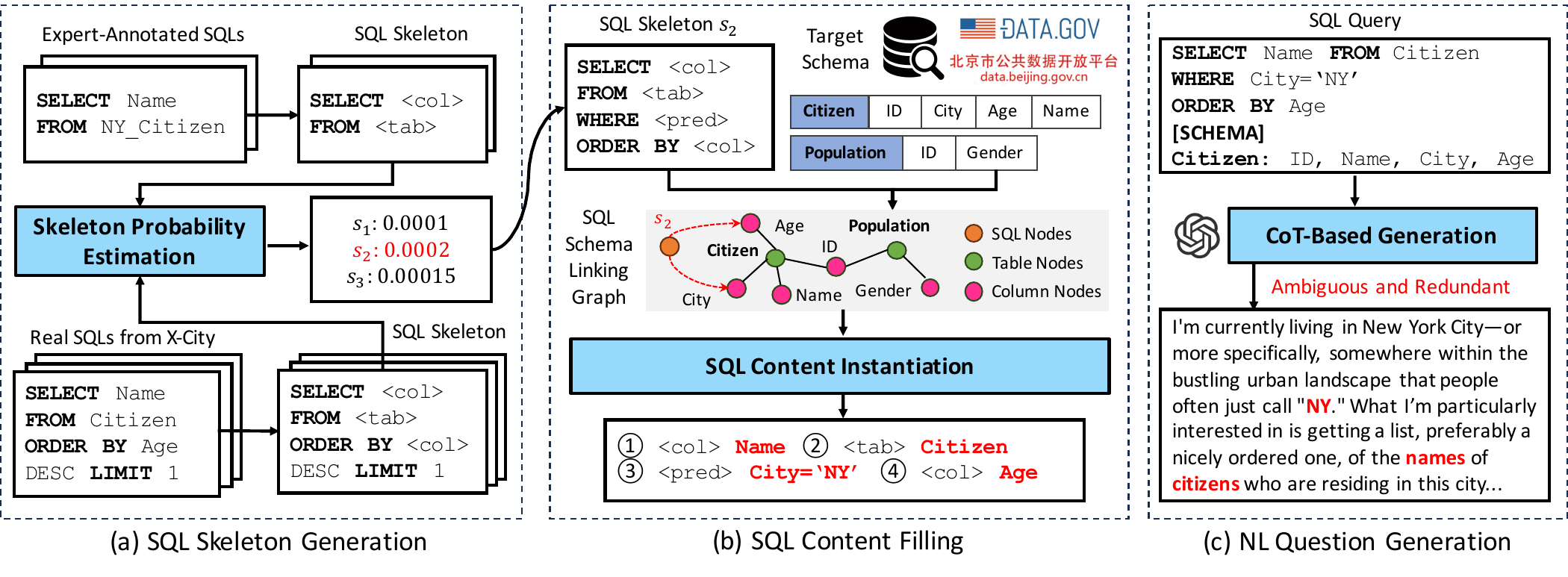}
    \vspace{-1em}
    \caption{Our three-step data synthesis pipeline for generating realistic NL--SQL \((Q, S)\) pairs.}
    \vspace{-1em}
    \label{fig:dataset2}
\end{figure*}

To protect privacy while retaining real-world structure, we construct a synthetic dataset from publicly available open-data portals, including the Beijing Municipal Open Data Platform~\cite{beijingdata} and the U.S.\ Government’s Open Data Portal~\cite{usdata}. These portals cover domains such as public services, healthcare, and transportation.
We collect raw files (CSV, XLSX, JSON), parse and clean them, and convert them into relational databases, resulting in 52 databases with 13{,}004 tables. Example~\ref{ex:synthetic-crossdb} illustrates a representative cross-database query in \sys-OpenData, where education and health datasets from different portals are joined via a shared geographic key.

\begin{example}[\blue{Synthetic Data: Cross-Database JOIN over Open Data Portals}]
\label{ex:synthetic-crossdb}
The query asks: \emph{``Join student performance data with community health statistics to examine the relationship between public health indicators and academic achievement across districts''}. 
Answering this query requires combining education and health datasets from different open-data portals via a shared geographic key. 

The corresponding SQL is:
\begin{verbatim}
SELECT 
  e."DistrictName",
  e."NumberofStudentsInNumerator",
  h."Community Area",
  h."Birth Rate",
  h."Mortality Rate"
FROM "State of Washington"."cte_perkins_concentrators
_2022_cohort" e
JOIN "City of Chicago"."public_health_statistics__
selected_public_health_indicators_by_chicago_community
_area__historical" h
  ON e."DistrictName" = h."Community Area"
  WHERE e."schoolyear" = '2022';
\end{verbatim}
\end{example}

However, these portals provide only raw datasets (databases, tables, and schemas), but neither executable SQL workloads nor paired NL questions. As a result, they cannot be directly used for open-domain \nlsql evaluation. We therefore develop a three-step synthesis pipeline (Figure~\ref{fig:dataset2}) to generate NL-SQL pairs \((Q, S)\): it preserves structural patterns of real SQL queries, populates query content using the public schemas, and generates corresponding NL questions via chain-of-thought promptings.
%
%

\etitle{Step 1: SQL Skeleton Generation.}
We introduce \emph{SQL skeletons}, i.e., {abstract templates that capture high-level structural patterns of SQL queries.
To approximate real-world usage, we estimate a \emph{skeleton distribution} from real queries and a small set of expert-designed SQL queries, and sample from this distribution to obtain representative skeletons (see Section~\ref{subsubsec:skeleton-gen}).

\etitle{Step 2: SQL Content Filling.}  
We instantiate each skeleton based on concrete schemas.
Databases are represented as a \emph{schema graph}, and a linking algorithm maps skeleton components to specific tables, columns, and predicates, ensuring semantic consistency and enabling cross-database composition (see Section~\ref{subsubsec:content-fill}).

\etitle{Step 3: NL Question Generation.}  
We generate NL questions using an LLM-based Chain-of-Thought (CoT) procedure conditioned on schema context and SQL semantics, which can include ambiguity and redundancy observed in user inputs (see Section~\ref{subsubsec:nl-gen}).

Using this pipeline, we construct a bilingual synthetic dataset with 13{,}000 \nlsql examples across 52 databases and 13{,}004 tables: 7{,}000 examples from the Beijing platform~\cite{beijingdata} and 6{,}000 from the U.S.\ portal~\cite{usdata}. 
These examples cover the four query categories introduced in Section~\ref{sec:pre} (single-DB lookups, multi-table joins, data-lake table retrieval, and cross-database compositions), allowing \sys-OpenData to reflect the open-domain query patterns observed in \sys-SmartCity.

\subsubsection{SQL Skeleton Generation}
\label{subsubsec:skeleton-gen}

We model the structural patterns of real SQL queries using \textbf{SQL skeletons}~\cite{DBLP:journals/pacmmod/GuF00JM023}. A skeleton represents the query structure using only SQL commands and operators, with placeholders such as \texttt{[tab]}, \texttt{[col]}, and \texttt{[val]} for schema- and value-specific content.
%
%
\begin{example}
Figure~\ref{fig:dataset2}(a) illustrates skeleton extraction: replacing all identifiers in a real SQL query with placeholders yields an SQL skeleton, e.g., \term{``SELECT [col] FROM [tab] ORDER BY [col] DESC LIMIT 1''}.
%
\end{example}
We extract SQL skeletons from two sources:
(1) existing \nlsql queries from Beijing, capturing generic open-domain patterns such as ambiguous questions, unspecified databases, and cross-database querying, and  
(2) a small set of expert-annotated SQL queries over the open-data schemas, encoding domain-specific logic from the Beijing and U.S.\ portals.

To model the \emph{distribution} of SQL structures, we follow prior work~\cite{xu2017sqlnetgeneratingstructuredqueries, yu2018syntaxsqlnetsyntaxtreenetworks} and decompose each SQL query into an abstract syntax tree (AST) represented as a sequence of context-free grammar (CFG) rules.
We estimate rule probabilities from frequency counts and combine rules from the two sources via a weighted mixture, yielding a global distribution over SQL structures.

Skeletons are generated top-down by sampling CFG rules from this distribution.
This ensures syntactic validity and aligns generated skeletons with prevalent query patterns in the target environment.
Finally, we select a diverse subset of skeletons: high-probability skeletons cover frequent patterns, while lower-probability ones capture realistic long-tail structures~\cite{yang2024synthesizingtexttosqldataweak, qin2024relationaldatabaseaugmentedlarge}.
Section~\ref{subsec:benchmark-statistics} reports the number of distinct skeletons observed in the final dataset.

\subsubsection{SQL Content Filling}
\label{subsubsec:content-fill}
Given the generated skeletons, we instantiate them into executable SQL queries over the target schemas (Figure~\ref{fig:dataset2}(b)). We represent each schema as an \emph{SQL-Schema Linking Graph (SSLG)}, whose nodes correspond to tables and columns and whose edges encode relationships such as primary-foreign key links~\cite{10711192, 10.1145/3062341.3062365}. The filling procedure consists of three stages:
%

\etitle{(1) Candidate Selection via SSLG.}
We construct the SSLG over the database schema. The graph encodes valid structural connections and is used to propose candidates when instantiating clauses such as \term{FROM} and \term{JOIN}.

\etitle{(2) Incremental Filling of Query Components.}
Guided by the skeleton (e.g., \term{SELECT}, \term{FROM}, \term{WHERE}, \term{GROUP BY}), we instantiate missing elements step by step.
The SSLG constrains choices to semantically valid tables, columns, and join paths, ensuring that generated queries comply with schema constraints and foreign-key structure.

\etitle{(3) LLM-Guided Predicate Generation.}
To produce realistic predicates and values, we employ an LLM conditioned on the skeleton and SSLG context~\cite{wang2023knowledgedrivencotexploringfaithful, qin2024relationaldatabaseaugmentedlarge}. For example, when instantiating a \term{WHERE} clause, the model generates filters (e.g., date ranges, regions, or thresholds) consistent with column types. The resulting SQL queries are executed to verify syntactic validity and non-empty outputs; invalid queries will be discarded or repaired in subsequent iterations.

\subsubsection{NL Question Generation}
\label{subsubsec:nl-gen}

Given a synthesized SQL query, this step generates an NL question that is consistent with the SQL semantics while exhibiting ambiguity and redundancy typical of open-domain inputs.
%
We use an LLM-based Chain-of-Thought (CoT) procedure with three phases:

\etitle{(1) SQL Semantic Interpretation.}
The model parses the SQL structure (e.g., \term{SELECT} targets, \term{JOIN} conditions, \term{WHERE} filters) together with schema metadata (table/column descriptions and relationships) to derive a natural-language description of the query semantics.

\etitle{(2) Query Intent Inference.}
The model maps the derived semantics to plausible user intents (e.g., monitoring service coverage, comparing districts, tracking trends).
%

\etitle{(3) Scenario-Conditioned Query Generation.}
The model generates NL questions by conditioning on the inferred intent and schema context, adding contextual details (e.g., time, location, user role) and allowing colloquial phrasing and implicit constraints, while preserving semantic equivalence to the SQL.


\subsubsection{Quality Control of the Synthesis Process}  
\label{sec:synthetic_validation}

Quality control is embedded in the synthesis pipeline. 
During NL generation, we use CoT prompts that explicitly ask the LLM to verify whether the generated query is consistent with the SQL semantics and schema context, and low-confidence outputs would be regenerated.
In addition, a lightweight LLM is used as a consistency checker: for each candidate NL-SQL pair, it scores semantic alignment with the associated schema, and only high-scoring pairs are retained.
%
We further conduct a stratified human evaluation on sampled synthetic examples (stratified by SQL complexity) to verify semantic correctness, fluency, and consistency. As reported in Table~\ref{tab:paper_human_eval_synthetic}, the synthetic NL-SQL pairs obtain average scores of 4.57/5 for semantic correctness and 4.61/5 for fluency, with strong inter-annotator agreement, providing evidence for the reliability of the synthesized data.


\subsection{Dataset Statistics of \sys}
\label{subsec:benchmark-statistics}
\begin{figure}[t!]
    \centering
    \includegraphics[width=\columnwidth]{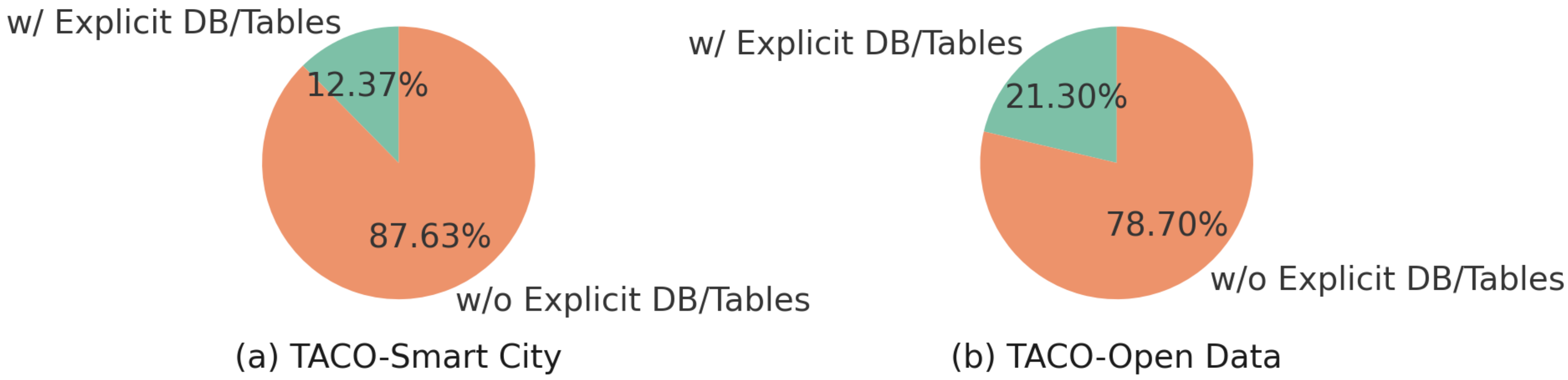}
    \caption{Proportion of NL questions with explicit DB/table mentions in \sys-SmartCity and \sys-OpenData.}
    \label{fig:table_linking_pie}
\end{figure}

\begin{figure}[t!]
    \centering 
    \includegraphics[width=\columnwidth]{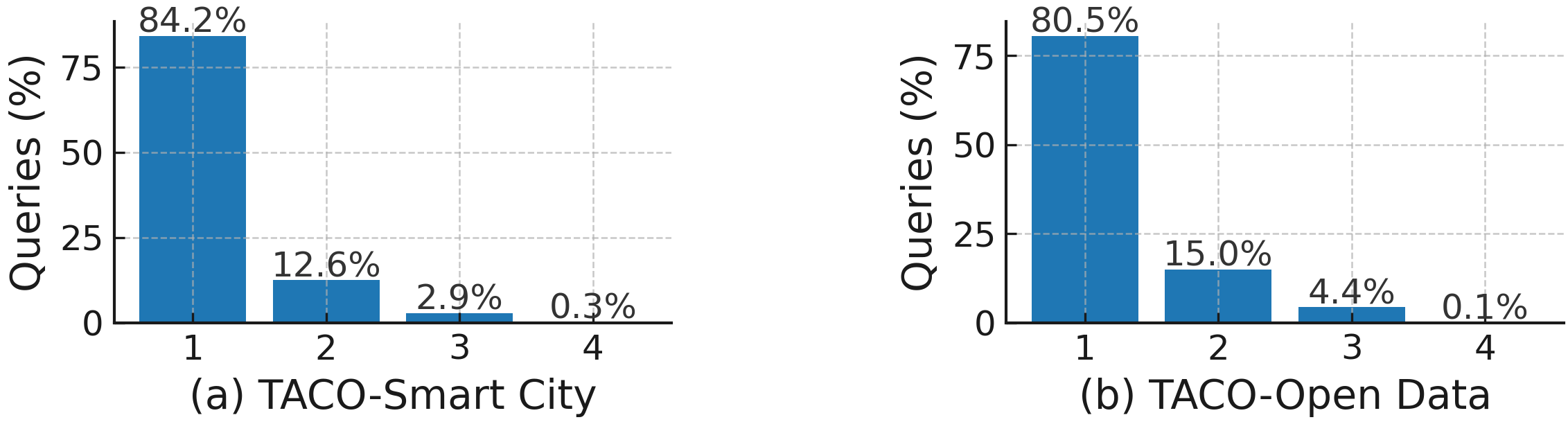}
    \caption{Number of databases involved in each query.}
    \label{fig:sql_number_distribution}
    \vspace{-1em}
\end{figure}


\begin{table}[t!]
  \centering
  \caption{\blue{Human evaluation of synthetic NL--SQL pairs.}}
  \vspace{-0.5em}
  \resizebox{\linewidth}{!}{%
  \begin{tabular}{lcccc}
    \toprule
    \textbf{Dataset} & \textbf{\#Samples} & \textbf{Correct} & \textbf{Minor Fix} & \textbf{Incorrect} \\
    \midrule
    TACO-OpenData-Beijing & 120 & 82\% & 13\% & 5\% \\
    TACO-OpenData-US      & 120 & 79\% & 15\% & 6\% \\
    \midrule
    Overall               & 240 & 80.5\% & 14\% & 5.5\% \\
    \bottomrule
  \end{tabular}
  }
  \vspace{0.4em}

  \resizebox{\linewidth}{!}{%
  \begin{tabular}{lccc}
    \toprule
    \textbf{Dataset} & \textbf{Fluency (1--5)} & \textbf{Realism (1--5)} & \textbf{Agreement $\kappa$} \\
    \midrule
    TACO-OpenData-Beijing & 4.3 & 4.1 & 0.78 \\
    TACO-OpenData-US      & 4.2 & 4.0 & 0.75 \\
    Overall               & 4.25 & 4.05 & 0.77 \\
    \bottomrule
  \end{tabular}
  }
  \label{tab:paper_human_eval_synthetic}
\end{table}

This section provides a quantitative characterization of TACO, focusing on (i) SQL/NL complexity, (ii) implicit schema references, (iii) cross-database usage, and (iv) structural diversity. Beyond reporting dataset scale, these analyses characterize the sources of difficulty in the benchmark and compare the synthetic \sys-OpenData with the real \sys-SmartCity subset, showing that the synthetic dataset preserves similar query patterns and structural properties.
%

\subsubsection{Statistics of \sys-SmartCity}

The real-world subset contains 113 tables and 1{,}500 user queries from 31 municipal departments. Key statistics are summarized below.
%

\etitle{(1) NL and SQL Length.}
Figure~\ref{fig:average_token_length} reports average token counts across benchmarks. \sys-SmartCity queries are significantly longer (SQL: 51.43 tokens; NL: 89.30 tokens) than \emph{Spider} and \emph{BIRD}, indicating increased linguistic redundancy and more complex query intents.

\etitle{(2) Implicit Schema References.}
As shown in Figure~\ref{fig:table_linking_pie}, 87.63\% of NL questions do \emph{not} explicitly mention any database or table, highlighting the need for table retrieval and schema linking.

\etitle{(3) Cross-Database Querying.}
Figure~\ref{fig:sql_number_distribution} shows that about 20\% of queries involve multiple databases. Many require multi-step pipelines with intermediate result reuse, increasing complexity relative to closed-domain benchmarks.




\subsubsection{Statistics of \sys-OpenData}

We evaluate whether \sys-OpenData exhibits characteristics comparable to \sys-SmartCity. The synthetic data contains 13{,}000 NL-SQL pairs across 52 databases with 13{,}004 tables.

\etitle{(1) Token Length and Redundancy.}
Figure~\ref{fig:average_token_length} shows that SQL and NL token-length distributions of the synthetic data are similar to those of the real queries, indicating comparable levels of redundancy.
%

\etitle{(2) Implicit Schema References.}
As shown in Figure~\ref{fig:table_linking_pie}, 78-87\% of \sys-OpenData queries do not explicitly mention a database or table, aligning with the \sys-SmartCity distribution.

\etitle{(3) Cross-Database Query Patterns.}
Figure~\ref{fig:sql_number_distribution} shows that the synthetic data includes frequent cross-database querying behaviors, with many queries combining information from 2-4 databases, consistent with the real subset.
%

\etitle{(4) SQL Structural Complexity.}
Figure~\ref{fig:revision_distribution} reports that synthetic SQL queries in \sys-OpenData are comparable to the real dataset in \sys-SmartCity in length, number of joins, and subquery depth, following heavy-tailed distributions. 
The prevalence of long queries, multi-hop joins, and nested subqueries exceeds that of benchmarks such as Spider and BIRD.
%

\etitle{(5) Diversity of SQL Skeletons.}
We count distinct SQL skeletons and find 2,317 unique skeletons in the synthetic dataset, suggesting that the workload is not limited to a small set of repeated structures.



\begin{figure*}[t!]
    \centering    \includegraphics[width=0.98\textwidth]{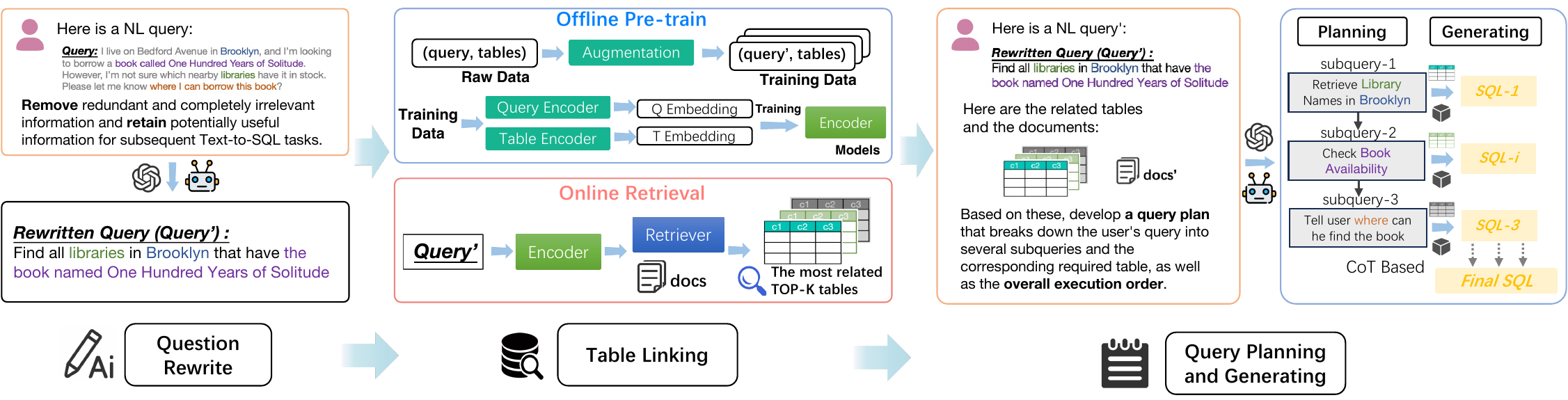}
    \caption{An overview of the \model framework for open-domain \nlsql.}
    \label{fig:our_framework}
    \vspace{-1em}
\end{figure*}

\etitle{Summary.}
\sys-OpenData closely matches \sys-SmartCity along multiple measurable dimensions while remaining public and reproducible. In particular, the two subsets exhibit similar distributions in SQL complexity, implicit schema linking, cross-database querying, and template diversity, indicating that the synthesis procedure preserves key workload characteristics of open-domain \nlsql.

%

\section{TACO-SQL FOR OPEN-DOMAIN TEXT-TO-SQL}
\label{sec:Framework}

To illustrate the use of \sys and evaluate the strengths and limitations of current approaches, we introduce \sys-SQL, an LLM-based framework for open-domain \nlsql, as shown in Figure~\ref{fig:our_framework}.
\sys-SQL is designed to address key challenges such as query ambiguity, large-scale table linking, and the complexity of cross-database query planning. The framework consists of three key components: (1) \emph{Question Rewriting}, (2) \emph{Table Linking}, and (3) \emph{Query Planning and Generating}, each addressing a specific aspect of open-domain \nlsql.

\stitle{Remarks.}
The primary contribution of this paper is the \sys benchmark and its analysis of open-domain \nlsql. \sys-SQL only serves as a reference baseline that exposes representative failure modes and supports modular replacement of components in future work.
%



\subsection{Question Rewriting}
In this component, we employ an LLM to rewrite noisy or an ambiguous user NL question into an explicit intent statement while preserving entities and constraints required for downstream table linking. The rewriting step normalizes ambiguous expressions, removes redundant conversational content, and makes implicit conditions explicit (e.g., identifiers, attributes, or target objects). 


For instance, the original query ``\emph{I need my employee records to finish a report. Please tell me where I can get my employee records. My employee ID is E12345, Thanks.}'' is rewritten as ``\emph{Find storage locations for employee records with ID E12345}''. 
The rewritten query explicitly exposes the target entity (employee records) and constraint (ID E12345), enabling more reliable table retrieval in the next stage.

\subsection{Table Linking}
This component identifies relevant tables from large, heterogeneous databases given the rewritten query, reducing the schema search space for downstream query planning and SQL generation~\cite{10.1145/3178876.3186067, zhang2020webtableextractionretrieval, wang2025dbcopilot, glass2025extractiveschemalinkingtexttosql}.
As illustrated in Figure~\ref{fig:our_framework}, table linking consists of two stages: an \emph{offline contrastive fine-tuning} phase that aligns query and schema representations, and an \emph{online retrieval} phase that retrieves candidate tables at inference time.
In the current implementation, table linking operates at the \emph{whole-query} level: the NL question is treated as a single unit to retrieve the top-$k$ tables from the entire data lake, while finer-grained decomposition is handled downstream in the query planning stage (Section~\ref{subsec:query-planning}).


\stitle{Offline Contrastive Fine-Tuning.}  
In the offline stage, we construct a labeled dataset $\mathcal{D} = \{(q_i, T_i^+)\}_{i=1}^{N}$, where each query $q_i$ is associated with a set of relevant tables $T_i^+$. Then, we adopt a dual-encoder architecture that embeds both the queries and table schema descriptions into a shared semantic space. Specifically, let $f_q(q)$ and $f_t(t)$ denote the query and table encoderS, respectively. The similarity between a query $q$ and a table $t$ is computed using cosine similarity:
\[
\text{sim}(q, t) = \frac{f_q(q) \cdot f_t(t)}{\|f_q(q)\| \, \|f_t(t)\|}.
\]
To ensure that matching query-table pairs are closely aligned, while non-matching pairs are separated, we optimize the encoders using the InfoNCE loss:
\[
\mathcal{L} = -\frac{1}{N}\sum_{i=1}^{N} \log \frac{\exp\left(\text{sim}(q_i, t_i^+)/\tau\right)}{\exp\left(\text{sim}(q_i, t_i^+)/\tau\right) + \sum_{t \in \mathcal{N}_i} \exp\left(\text{sim}(q_i, t)/\tau\right)},
\]
where $\tau$ is a temperature parameter and $\mathcal{N}_i$ denotes a set of negative tables for query $q_i$. 
The objective increases the similarity of positive query–table pairs relative to negatives, encouraging alignment between natural-language queries and schema descriptions~\cite{Zhang_2019, xiao2022distillvqlearningretrievaloriented}.
%

\stitle{Online Retrieval.}  
In the online stage, the fine-tuned dual-encoder is employed to process incoming user queries. For a query $q$, we compute its embedding $f_q(q)$, while the embeddings of candidate tables $\{f_t(t_j)\}_{j=1}^{M}$ are pre-computed and indexed for efficiency. 
We then compute cosine similarity $\text{sim}(q, t_j)$ between the query and each table, and rank the candidate tables. The top-$k$ relevant tables are selected as:
\[
\mathcal{T}^* = \{t_j \mid \text{rank}(\text{sim}(q, t_j)) \leq k\}.
\]

To improve recall under heterogeneous naming and phrasing, we apply two augmentation strategies~\cite{pmlr-v195-parulekar23a}. \emph{Query variants} are generated via synonym substitution and paraphrasing, and \emph{schema augmentation} expands abbreviations and enriches table descriptions~\cite{rebuffi2021dataaugmentationimproverobustness, chen2022adversarialtrainingimprovingmodel, yu2021score}.
%
Section~\ref{sec:Experiments} reports an ablation comparing whole-query linking with a subquery-aware variant and a hybrid strategy, measured by gold-table recall and subquery coverage.

\subsection{Query Planning and Generating}  
\label{subsec:query-planning}
Queries in open-domain \nlsql may span multiple databases and contain implicit dependencies across sub-queries. To handle such complexity, we transform the rewritten question $q$, the retrieved tables $\mathcal{T}^*$ and associated schema text $\mathcal{D}^*$ (e.g., table descriptions) into a structured execution plan $\mathcal{P}^*$ used for SQL generation.

\stitle{Structured Execution Plan.} 
We construct a prompt containing $q$, descriptions of the tables in $\mathcal{T}^*$, and key schema information, and provide the prompt to an LLM to produce an execution plan:
\[
\mathcal{P}^* = f_{\text{plan}}(q, \mathcal{T}^*, \mathcal{D}^*),
\]
where $f_{\text{plan}}(\cdot)$ denotes the planning function implemented via LLM prompting. The output $\mathcal{P}^*$ is an ordered list of sub-queries $\{p_1, p_2, \ldots, p_n\}$, each of which is associated with its corresponding table(s) and operation, specifying the execution order.
This prompting-based strategy leverages the LLM’s reasoning capabilities to infer sub-query dependencies, eliminating the need for manually crafted dependency graphs. The resulting plan ensures that inter-table operations are logically and contextually sequenced.
Each subquery in $\mathcal{P}^*$ corresponds to one SQL statement, and the resulting plan forms an \emph{SQL-only} pipeline whose intermediate results are combined within the database engine using standard SQL operators, consistent with the problem formulation in Section~\ref{subsec:problem formulation}.
%

\stitle{SQL Generation via Step-by-Step Reasoning.}  
Unlike direct NL-to-SQL generation, we leverage a step-by-step method to generate intermediate steps that guide the final SQL generation. Given the execution plan $\mathcal{P}^*$, relevant tables $\mathcal{T}^*$, and schema $\mathcal{D}^*$, we decompose the NL question into several subqueries as a sequence of reasoning steps $C = \{c_1, c_2, \ldots, c_n\}$, along with corresponding table descriptions and key schema details. Then, we apply an existing \nlsql model to generate an SQL query:
\[
S_{\text{draft}} = g(C, \mathcal{P}^*, \mathcal{T}^*, \mathcal{D}^*),
\]
where $g(\cdot)$ denotes the SQL generation model.

\stitle{Iterative Refinement and Error Correction.}
We employ an iterative refinement procedure driven by automatic error detection and feedback. Specifically, intermediate and final SQL queries are executed on the underlying databases, and execution errors are returned to the LLM to revise the query. All refinement and result computation are performed entirely within the database engine using SQL, without external scripts. The iteration terminates once an executable SQL query is produced, which is then evaluated against the gold SQL in \sys.


\section{EXPERIMENTS}
\label{sec:Experiments}

In this section, we evaluate the \sys benchmark and the proposed \model pipeline through three key sets of experiments.
\begin{itemize}[leftmargin=*]
    \item \textbf{Exp-1} evaluates existing \nlsql approaches on \sys without additional preprocessing, measuring performance under the open-domain setting.
    \item \textbf{Exp-2} analyzes how query-level factors, such as NL length, schema size, and cross-database complexity, affect performance.
    \item \textbf{Exp-3} examines the contribution of each component in \model through ablation studies, e.g., \emph{Question Rewriting}, \emph{Table Linking}, and \emph{Query Planning}.
\end{itemize}

\subsection{Experimental Setup}
This section first presents the experimental setup to evaluate \nlsql approaches on \sys. 

\stitle{Baselines.}
We evaluate a diverse set of \nlsql approaches under the open-domain setting.

\etitle{(1) Base LLMs.}
General-purpose LLMs such as GPT-4, GPT-4o, GPT-o1, Llama-3, and DeepSeek-v3 are evaluated in the zero-shot setting.
This setting measures model performance when applied directly to large and heterogeneous schemas without task-specific training.


\etitle{(2) LLM-Based Methods.}
Approaches such as \emph{DIN-SQL}~\cite{pourreza2023dinsqldecomposedincontextlearning} and \emph{MAC-SQL}~\cite{wang2024macsqlmultiagentcollaborativeframework} rely on structured prompting or decomposition strategies for reasoning over complex queries.

\etitle{(3) SFT-Based Methods.}
Models including \emph{CodeS-33B}~\cite{10.1145/3654930} and \emph{Qwen2.5-Coder-32B} are fine-tuned on SQL and code corpora.
This setting evaluates the effect of supervised fine-tuning when the target database is not specified.
%

\etitle{(4) Hybrid Methods.}
Systems such as \emph{CHESS}~\cite{talaei2024chesscontextualharnessingefficient} and \emph{ZeroNL2SQL}~\cite{10.14778/3681954.3681960} combine LLM reasoning with schema linking or iterative refinement, representing practical strategies for \nlsql.

\stitle{Method Configurations.} 
All models are evaluated under a unified and reproducible setup, including consistent prompting templates, schema-visibility settings, and decoding configurations (subject to model-specific context limits). As \sys is released as an evaluation-only benchmark, all examples are treated as a held-out test set. To support reproducibility, we provide prompt templates, configuration files, and evaluation scripts in our GitHub repository.

\stitle{Schema Visibility Settings.}
Models receive the raw user query together with the full schema collection (database–table–column hierarchy), reflecting the open-domain setting and exposing models to large heterogeneous schema spaces. In some cases, context-length limits lead to partial truncation and affect performance. 
Moreover, to ablate the impact of table retrieval, Exp-3 additionally reports results where models are provided with the gold relevant tables, so that downstream SQL reasoning can be evaluated independently of the table-linking stage.

%

\stitle{Evaluation Metric.}
We adopt Execution Accuracy (EX)~\cite{yu2019spiderlargescalehumanlabeleddataset}, following prior benchmarks such as BIRD~\cite{li2024can}. EX measures whether the result returned by the predicted SQL matches the result of the ground-truth SQL.


\begin{table}[t]
  \centering
  \caption{Baseline Performance on the \sys benchmark, reported as Execution Accuracy (EX) in \%. The best results in each dataset are shown in bold.
  A hyphen (``-'') indicates that the dataset cannot be evaluated by closed-source models due to privacy restrictions, and therefore no result is reported.}

  \vspace{-0.5em}
\resizebox{\linewidth}{!}{
    \begin{tabular}{|c||c|c|c|}
      \hline
      \multirow{2}{*}{\textbf{Baselines}} & \multicolumn{3}{c|}{\textbf{TACO Benchmarks}} \\ \cline{2-4}
      & \textbf{\sys-SmartCity} & \textbf{\sys-Beijing} & \textbf{\sys-US} \\ 
      \hline\hline
      GPT-4                 & - & 11.50 & 12.84 \\ \hline
      GPT-4o                & - & 12.06 & 13.16 \\ \hline
      GPT-o1                & - & \textbf{14.18} & \textbf{15.32} \\ \hline
      DeepSeek-v3           &  - &  11.30 &  12.28 \\ \hline
      Llama3-70B            &  5.33 &  6.66 &  8.16 \\ \hline\hline
      DIN-SQL (GPT-4o)        &  - &  7.48 &  8.02 \\ \hline
      MAC-SQL (GPT-4o)        &  - &  6.90 &  7.64 \\ \hline\hline
      CodeS-33B             &  \textbf{8.47} &  9.02 &  9.96 \\ \hline
      Qwen2.5-Coder-32B     &  5.91 &  6.52 &  7.28 \\ \hline\hline
      CHESS                 &  - & 10.10 & 11.84 \\ \hline
      Zero-NL2SQL           &  - &  7.62 &  8.26 \\ \hline
    \end{tabular}
    \vspace{-2em}
}
  \label{tbl:vs_other_prep_method_origin}
\end{table}

\subsection{Exp-1: Baseline Performance on \sys}
\label{sec:rq1_baseline}

This experiment evaluates how existing \nlsql approaches, which are largely developed for \emph{closed-domain} settings, perform under the open-domain conditions of \sys.  
All baselines are evaluated \emph{without} auxiliary preprocessing: models receive only the raw user question and the full collection of database schemas, with no question rewriting, table linking, or planning modules.
This configuration measures performance when \nlsql approaches are directly applied to the open-domain setting. 

Table~\ref{tbl:vs_other_prep_method_origin} reports execution accuracy across a broad range of model families.
%
All models achieve low execution accuracy ($6-15\%$). This result indicates that \sys presents challenges even for strong LLMs. The performance gap is consistent with factors such as ambiguous user intent, large schema search spaces, implicit table references, and cross-database reasoning, which are less emphasized in traditional benchmarks.

Among all baselines, GPT-based models (e.g., GPT-4o and GPT-o1) and  DeepSeek-v3 achieve better results. For example, GPT-o1 reaches $14-15\%$ on \sys-Beijing and \sys-US.  
The improvement over other baselines is consistent with the broader pretraining of general-purpose LLMs and their handling of long queries and large schema contexts.
%
SFT-based models such as \emph{CodeS-33B} and \emph{Qwen2.5-Coder-32B} achieve moderate performance ($8-10\%$), suggesting that SQL-focused fine-tuning alone does not address open-domain ambiguity or table-retrieval challenges.
Task-specific prompting methods such as \emph{DIN-SQL} and \emph{MAC-SQL} perform comparably, but their performance decreases relative to closed-domain benchmarks.

\stitle{Finding 1.}
Current \nlsql approaches show limited performance under the open-domain setting.
The low execution accuracy demonstrates that \sys reveals limitations in both base LLMs and SQL-tuned models, particularly for queries without explicit table references and those requiring multi-step cross-database reasoning. 
These observations motivate the use of modular components, such as question rewriting, table linking, and structured planning, to addresss the challenges of open-domain \nlsql. 


\begin{figure*}[t]
    \centering     
    \includegraphics[width=1\textwidth]{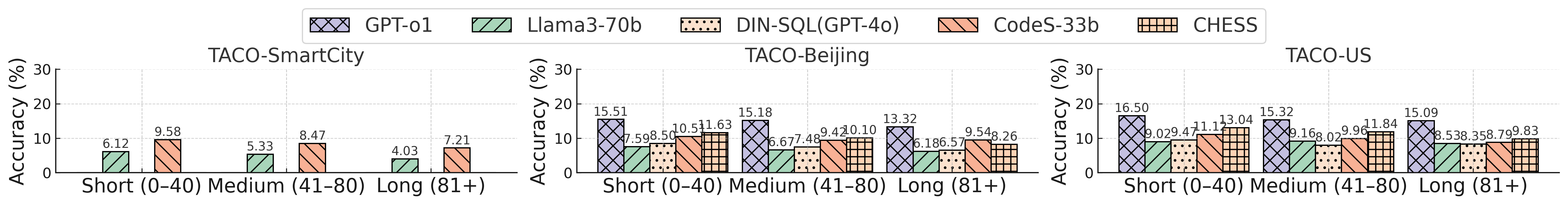}
    \caption{Effect of NL Query Length (Execution accuracy decreases with longer queries).}
    \label{fig:a}
\end{figure*}

\begin{figure*}[t]
    \centering 
    \includegraphics[width=1\textwidth]{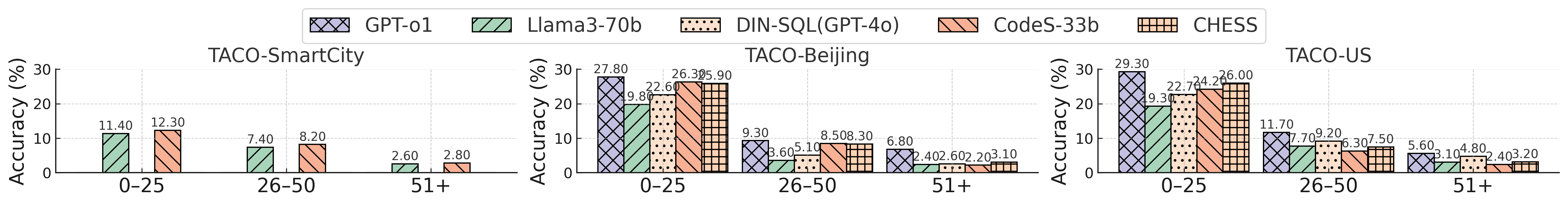}
    \caption{Effect of Schema Size (Execution accuracy decreases as the number of tables increases).}
    \label{fig:b}
\end{figure*}

\begin{figure*}[t]
    \centering
    \includegraphics[width=1\textwidth]{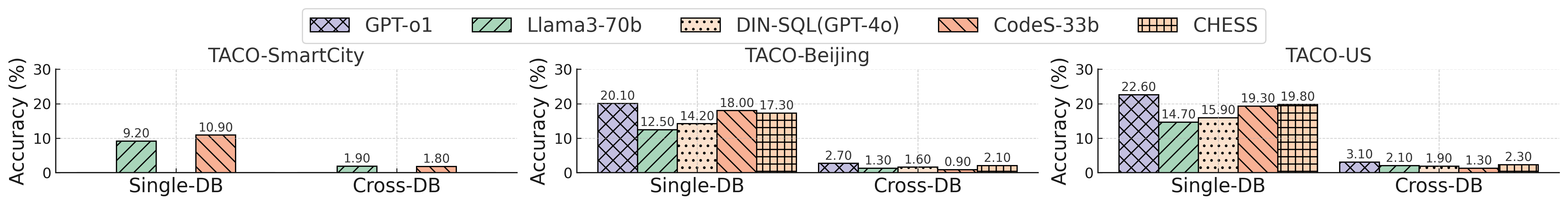}
    \caption{Effect of Cross-Database Complexity (Multi-database queries show lower accuracy than single-database queries).
    %
    }
    \label{fig:c}
\end{figure*}

\begin{table*}[t]
  \centering
  \caption{
    Model Performance on the TACO Benchmark Under Different Settings.
    Best results for each model are bolded. Origin: raw query with full schemas. QR: Question Rewriting. TL: gold relevant tables (no table retrieval). QR+TL: Question Rewriting with table linking. QR+TL+QP (\sys-SQL): adds Query Planning. A hyphen (``–'') indicates results are unavailable for closed-source models due to privacy restrictions.
%
}
  \resizebox{\linewidth}{!}{
    \begin{tabular}{|c||c|c|c|c|c||c|c|c|c|c||c|c|c|c|c|}
      \hline
      \multirow{2}{*}{\textbf{Models}} 
        & \multicolumn{5}{c||}{\textbf{\sys-SmartCity}} 
        & \multicolumn{5}{c||}{\textbf{\sys-Beijing}} 
        & \multicolumn{5}{c|}{\textbf{\sys-US}}  \\ \cline{2-16} 
        & \textbf{Origin} & \textbf{QR} & \textbf{TL} & \textbf{QR+TL} & \makecell[c]{\textbf{QR+TL+QP}\\(\textbf{\sys-SQL})}  
        & \textbf{Origin} & \textbf{QR} & \textbf{TL} & \textbf{QR+TL} & \makecell[c]{\textbf{QR+TL+QP}\\(\textbf{\sys-SQL})}  
        & \textbf{Origin} & \textbf{QR} & \textbf{TL} & \textbf{QR+TL} & \makecell[c]{\textbf{QR+TL+QP}\\(\textbf{\sys-SQL})}  \\ \hline\hline

      GPT-4                   
        & -     & -     & -     & -     & -     
        & 11.50 & 14.38 & \blue{21.72}    & 25.52 & \textbf{26.80} 
        & 12.84 & 16.92 & \blue{25.44}    & 28.36 & \textbf{30.04} \\ \hline

      GPT-4o                  
        & -     & -     & -     & -     & -     
        & 12.06 & 16.42 & \blue{26.44}    & 28.60 & \textbf{28.94} 
        & 13.16 & 16.28 & \blue{30.08}    & \textbf{31.76} & 30.04 \\ \hline

      GPT-o1                  
        & -     & -     & -     & -     & -     
        & 14.18 & 18.58 & \blue{29.16}    & 32.74 & \textbf{34.92} 
        & 15.32 & 22.06 & \blue{32.90}    & 33.24 & \textbf{35.18} \\ \hline

      DeepSeek-v3             
        & -     & -     & -     & -     & -     
        & 11.30 & 11.28 & \blue{19.22}    & 21.74 & \textbf{22.98} 
        & 12.28 & 13.08 & \blue{24.06}    & \textbf{25.50} & 23.66 \\ \hline

      Llama3-70B              
        &  5.33 &  6.54 & \blue{14.56}    & 19.53 & \textbf{19.60} 
        &  6.66 &  8.53 & \blue{17.78}    & 21.26 & \textbf{22.73} 
        &  8.16 &  9.60 & \blue{19.40}    & \textbf{20.58} & 19.67 \\ \hline\hline

      DIN-SQL(GPT-4o)          
        & -     & -     & -     & -     & -     
        &  7.48 &  9.92 & \blue{18.46}    & \textbf{21.76} & 20.52 
        &  8.02 & 11.38 & \blue{24.12}    & 25.82 & \textbf{26.40} \\ \hline

      MAC-SQL(GPT-4o)          
        & -     & -     & -     & -     & -     
        &  6.90 &  7.88 & \blue{15.06}    & 18.34 & \textbf{19.20} 
        &  7.64 &  7.92 & \blue{17.88}    & 19.60 & \textbf{19.88} \\ \hline\hline

      CodeS-33B               
        &  8.47 & 10.12 & \blue{24.86}    & 26.22 & \textbf{28.18} 
        &  9.02 & 13.56 & \blue{25.94}    & 28.32 & \textbf{28.78} 
        &  9.96 & 11.50 & \blue{27.38}    & \textbf{29.18} & 27.66 \\ \hline

      Qwen2.5-Coder-32B       
        &  5.91 &  7.65 & \blue{18.32}    & \textbf{22.83} & 19.79 
        &  6.52 &  8.52 & \blue{19.74}    & \textbf{22.06} & 20.34 
        &  7.28 &  9.40 & \blue{15.20}    & 18.14 & \textbf{20.38} \\ \hline\hline

      CHESS                   
        & -     & -     & -     & -     & -     
        & 10.10 & 10.02 & \blue{24.94}    & 28.30 & \textbf{30.22} 
        & 11.84 & 13.50 & \blue{27.12}    & 31.44 & \textbf{32.98} \\ \hline

      Zero-NL2SQL             
        & -     & -     & -     & -     & -     
        &  7.62 &  8.50 & \blue{23.88}    & 26.82 & \textbf{27.46} 
        &  8.26 &  9.10 & \blue{23.04}    & \textbf{25.74} & 25.36 \\ \hline

    \end{tabular}
  }
  \label{tbl:vs_other_prep_method}
\end{table*}

\subsection{Exp-2: Impact Factors on Open-Domain Text-to-SQL Performance}
\label{sec:rq2_bucket}

In this experiment, we analyze how the factors, {query length}, {schema size} (number of tables), and {cross-database complexity}, affect model performance on \sys.
The goal is to analyze performance at a finer granularity than aggregated EX scores and examine which properties of open-domain queries are associated with performance variation, complementing the dataset statistics in Section~\ref{subsec:benchmark-statistics}. Figures~\ref{fig:a}, \ref{fig:b} and \ref{fig:c} report execution accuracy bucketed by these factors, and Table~\ref{tbl:vs_other_prep_method} lists representative results for selected models.
We have the following observations.


\stitle{Effect of Query Length.}  
Figure~\ref{fig:a} shows that EX generally decreases as query length increases. On \sys-Beijing, for example, GPT-o1 achieves around 15\% EX for short queries (0-40 tokens), dropping to roughly 13\% for long queries (81+ tokens). A similar pattern appears on \sys-US for models such as CodeS-33B and Qwen2.5-Coder-32B. This trend is consistent with longer queries containing more redundancy and implicit conditions, which complicate NL-to-SQL mapping.

\stitle{Effect of Schema Size (\#Tables).}  
Figure~\ref{fig:b} reports the effect of schema size on performance.
On \sys-Beijing, GPT-o1 achieves over 25\% EX when the relevant schema contains 0-25 tables, but the EX drops below 10\% when more than 50 tables are present. Similar patterns hold across models and on \sys-US.
The results are consistent with the need to identify relevant tables before SQL generation.

\stitle{Effect of Cross-Database Complexity.}  
Figure~\ref{fig:c} shows lower performance on cross-database queries. On \sys-US, GPT-o1 exceeds 20\% EX on single-database questions but drops to only a few percent for cross-database ones, with similar gaps across baselines. Cross-database queries require coordinating multiple query steps and maintaining consistency across databases, which increases the complexity of the generation process.


\stitle{Finding 2.}
\nlsql performance decreases with longer queries, larger schema sizes, and cross-database complexity. Together with the structural statistics in Section~\ref{subsec:benchmark-statistics}, these results indicate that \sys captures the key challenges of open-domain \nlsql.


\subsection{Exp-3: The \sys-SQL Framework}
\label{sec:rq4_framework}

We evaluate the \sys-SQL framework by measuring the incremental contribution of its components on \sys.



Specifically, we evaluate the framework under five progressively stronger configurations:
(1) \textbf{Origin}: Direct SQL generation from the raw query with access to the full schema.
(2) \textbf{QR}: Applying Question Rewriting on top of \textbf{Origin}.
(3) \textbf{TL}: Providing the oracle set of gold relevant tables, ablating the table-retrieval stage and evaluating downstream SQL generation.
%
(4) \textbf{QR+TL}: Applying retrieval-based Table Linking after Question Rewriting.
(5) \textbf{\sys-SQL = QR+TL+QP}: Adding Query Planning to the QR+TL configuration.

\begin{table}[t]
  \centering
  \caption{Cross-Database Performance With vs. Without Query Planning (QP) (EX \%).}
\resizebox{\linewidth}{!}{
    \begin{tabular}{|c||cc||cc||cc|}
      \hline
      \multirow{2}{*}{\textbf{Models}} 
      & \multicolumn{2}{c||}{\textbf{\sys-SmartCity}} 
      & \multicolumn{2}{c||}{\textbf{\sys-Beijing}} 
      & \multicolumn{2}{c|}{\textbf{\sys-US}} \\ \cline{2-7}
      & \textbf{w/o QP} & \textbf{w/ QP} 
      & \textbf{w/o QP} & \textbf{w/ QP} 
      & \textbf{w/o QP} & \textbf{w/ QP} \\ \hline\hline

      GPT-4                   
      & - & -
      & 5.52 & \textbf{11.38}
      & 6.36 & \textbf{13.94} \\ \hline

      GPT-4o                  
      & - & -
      & 8.60 & \textbf{12.30}
      & 8.76 & \textbf{14.06} \\ \hline

      GPT-o1                  
      & - & -
      & 10.74 & \textbf{16.12}
      & 12.24 & \textbf{18.54} \\ \hline

      DeepSeek-v3             
      & - & -
      & 7.74 & \textbf{11.50}
      & 7.62 & \textbf{12.68} \\ \hline

      Llama3-70B              
      & 3.53 & \textbf{7.10}
      & 4.26 & \textbf{8.38}
      & 6.58 & \textbf{9.86} \\ \hline\hline

      DIN-SQL(GPT-4o)          
      & - & -
      & 7.12 & \textbf{10.04}
      & 8.82 & \textbf{13.94} \\ \hline

      MAC-SQL(GPT-4o)          
      & - & -
      & 6.34 & \textbf{9.76}
      & 8.66 & \textbf{10.02} \\ \hline\hline

      CodeS-33B               
      & 5.28 & \textbf{10.10}
      & 6.32 & \textbf{9.24}
      & 7.18 & \textbf{11.52} \\ \hline

      Qwen2.5-Coder-32B       
      & 4.83 & \textbf{9.47}
      & 6.06 & \textbf{8.28}
      & 7.14 & \textbf{9.50} \\ \hline\hline

      CHESS                   
      & - & -
      & 8.38 & \textbf{12.60}
      & 9.44 & \textbf{13.78} \\ \hline

      Zero-NL2SQL             
      & - & -
      & 6.82 & \textbf{10.50}
      & 7.74 & \textbf{11.12} \\ \hline
    \end{tabular}
}
  \label{tbl:cross_db_qp_impact}
\end{table}

\stitle{Overall Results.}
Table~\ref{tbl:vs_other_prep_method} summarizes results across the three datasets. We have the following observations.

\etitle{(1) Effect of Table Linking (QR+TL).}
Across models, applying Table Linking after Question Rewriting yields the largest incremental performance gain. For example, GPT-o1 increases from 14.18\% (\sys-Beijing, Origin) to 32.74\% (QR+TL), more than doubling EX. Similar gains appear for CodeS-33B, CHESS, and Qwen2.5-Coder-32B. These results indicate that providing retrieved tables instead of the full schema could significantly improves performance in the open-domain \nlsql setting.


\etitle{(2) Effect of Question Rewriting (QR).}
Applying Question Rewriting alone produces modest but consistent improvements. For example, GPT-4 increases from 11.50\% to 14.38\% on \sys-Beijing, with similar patterns across other models. QR clarifies query intent and normalizes expressions, but does not change the set of candidate tables provided for SQL generation, and therefore yields smaller gains than configurations that include Table Linking.


\etitle{(3) Evaluation with Gold Tables.}
Providing the gold relevant tables improves performance but remains below the QR+TL configuration. For example, GPT-4o achieves 26.44\% under TL versus 28.60\% under QR+TL on \sys-Beijing. The performance gap suggests that errors also arise in downstream SQL generation, such as join construction and subquery formulation.
%

\etitle{(4) Effect of Query Planning (QP).}
Adding Query Planning achieves additional improvements, though smaller than those from Table Linking. For example, the EX of GPT-o1 increases from 32.74\% (QR+TL) to 34.92\% (QR+TL+QP).
Moreover, the improvement is larger on cross-database queries. As shown in Table~\ref{tbl:cross_db_qp_impact}, GPT-o1 increases from 10.74\% to 16.12\% on \sys-Beijing for cross-database queries, and CodeS-33B from 5.28\% to 10.10\%. 
This pattern is consistent with settings where multiple query steps must be coordinated across databases.
QP mainly affects multi-step SQL composition. Without QP, models tend to express multiple operations in a single query, leading to incorrect joins, missing intermediate filtering, or misplaced aggregation. The structured plan decomposes the task into ordered subqueries, separating table selection, filtering, and aggregation. This is more useful in cross-database queries, where intermediate results should be produced and reused across databases rather than represented as a single flat query.


\etitle{(5) Impact Across Model Families.}
General-purpose LLMs such as GPT-o1 and GPT-4o achieve the strongest overall performance. However, fine-tuned models (e.g., CodeS-33B) and hybrid systems (e.g., CHESS) benefit more from the \sys-SQL pipeline, suggesting that SQL-oriented training and structured reasoning components remain particularly helpful under open-domain conditions.

%

\begin{figure}[t!]
    \centering
    \includegraphics[width=1.0\linewidth]{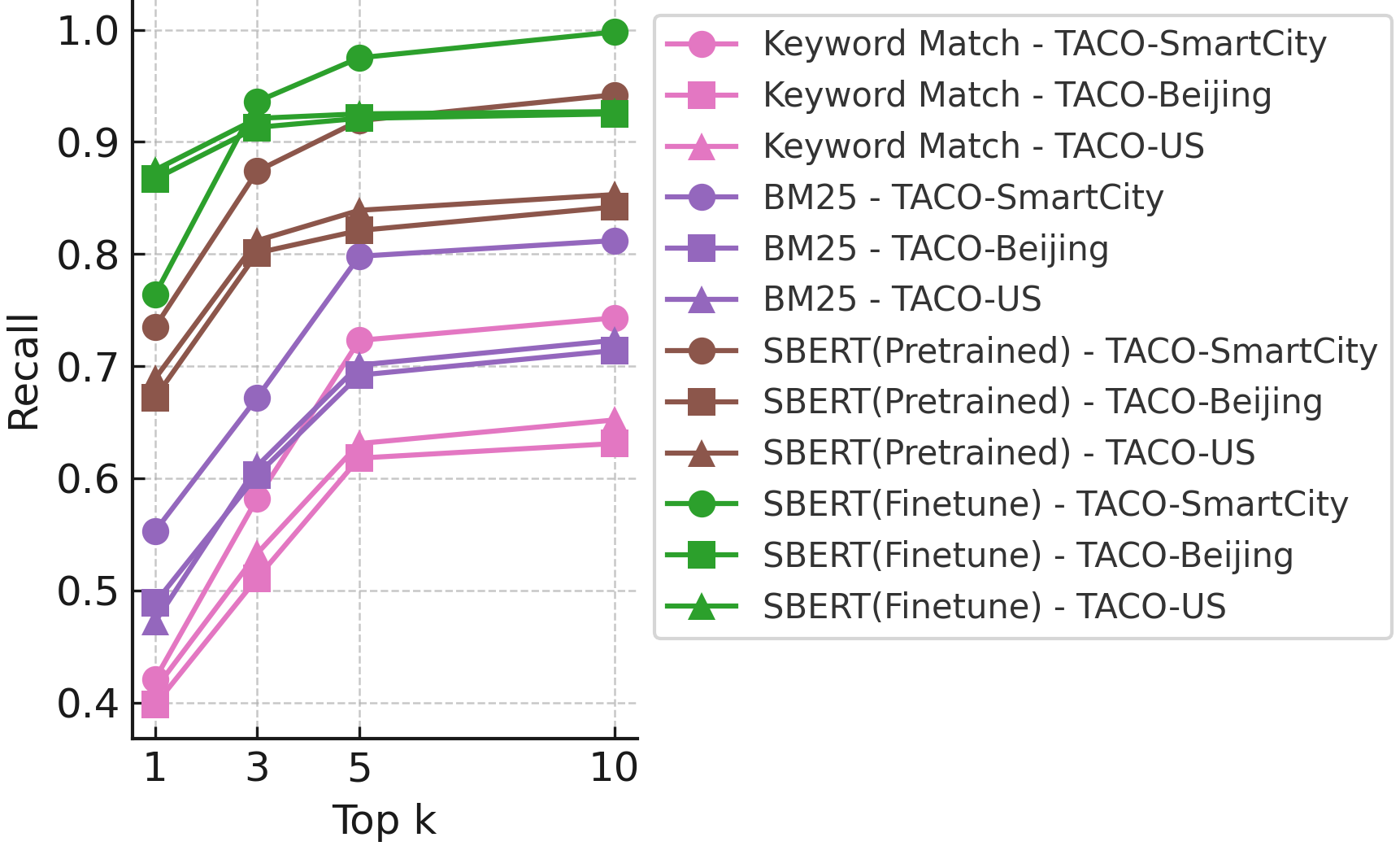}
    \caption{
    Evaluation on table linking performance under varying top-$k$
    (Larger $k$ increases recall but lower precision; $k=5$ provides a balanced setting).
    }
    \label{fig:table_linking_results}
\end{figure}

\stitle{Effect of Table Linking Design.}
We compare four table-linking methods, keyword matching, BM25, SBERT (pretrained), and SBERT (fine-tuned), using Recall@$k$ as the metric.
As shown in Figure~\ref{fig:table_linking_results}, increasing $k$ from 1 to 5 improves recall (by 5\%–10\%), whereas further increases yield only marginal gains.
Based on this trade-off between coverage and retrieval noise, we set $k=5$ for all subsequent experiments.


We also implement an alternative \textbf{subquery-level linking} strategy and evaluate it alongside our default whole-query approach. As shown in Table~\ref{tab:add_linking}, the subquery-based method achieves performance comparable to whole-query linking, with only marginal differences across evaluation metrics.
These results suggest that, under the open-domain conditions of \sys, the full query already provides sufficiently informative signals for table retrieval, and finer-grained decomposition yields limited additional benefit.


\begin{table}[t]
\centering
\small
\caption{
Whole-Query vs. Subquery Table Linking (200 cross-table/database queries).
}
\label{tab:add_linking}

\resizebox{0.95\linewidth}{!}{
\renewcommand{\arraystretch}{1.20} %
\begin{tabular}{|c||c||c|}
\hline
\textbf{Strategy} & 
\textbf{Recall@k} & 
\textbf{Subquery Cov.} \\
\hline\hline

Whole-query linking &
93.4\% &
91.2\% \\
\hline

Sub-query linking &
\textbf{94.2\%} &
\textbf{92.6\%} \\ 
\hline

%

\end{tabular}
}
\vspace{-1em}
\end{table}

\stitle{Finding 3.}
The results indicate that solving \sys requires three core capabilities: (i) NL query disambiguation (QR), (ii) table linking over large heterogeneous databases (TL), and (iii) multi-step and cross-database SQL planning (QP).
The \sys-SQL framework consistently improves performance across model types, supporting its modular design and highlighting practical directions for future open-domain \nlsql systems.

\section{CONCLUSION AND FUTURE WORK}
\label{sec:Conclusion_and_Future_Work}

In this paper, we have introduced \sys, a benchmark for evaluating open-domain \nlsql systems.
\sys consists of two datasets: \sys-SmartCity, which contains 1,500 real NL-SQL pairs curated through multi-stage expert annotation, and \sys-OpenData, which provides 13,000 synthetic examples generated by a data synthesis pipeline that preserves real-world structural and semantic complexity.
We have also presented \sys-SQL, an LLM-based baseline that integrates question rewriting, table linking, and query planning.
Our experiments show a substantial performance gap to gold SQL, highlighting the difficulty of open-domain \nlsql and establishing \sys as a challenging benchmark.

For future work, this benchmark raises several broader research directions.
Open-domain \nlsql requires evaluating not only final SQL correctness but also intermediate behaviors, including table discovery, ambiguity resolution, and multi-step planning. This motivates evaluation metrics beyond execution accuracy. In addition, real deployments operate over evolving data lakes where schemas and values change over time, and benchmarks that model dynamic environments can better evaluate robustness under schema drift. Finally, practical systems increasingly behave as agentic LLMs that interact with users and databases during problem solving. Handling underspecified intent may require clarification, feedback, and iterative querying. Thus, an interesting direction is to develop evaluation settings that measure interactive and multi-step reasoning rather than single-shot SQL generation.

\section*{ACKNOWLEDGMENTS}
This work was partially supported by the National Natural Science Foundation of China (Grant Nos. 62436010, 62441230, and 62402409) and the Scientific Research Innovation Capability Support Project for Young Faculty (Grant No. SRICSPYF-ZY2025001).
We also thank the Beijing Big Data Centre for their collaboration and for providing real-world data and application scenarios.

\bibliographystyle{ACM-Reference-Format}
\bibliography{refs/taco}


\end{document}